\documentclass{article}

\usepackage[preprint]{Formatting_Instructions_For_NeurIPS_2026/neurips_2026}

\usepackage[utf8]{inputenc}
\usepackage[T1]{fontenc}
\usepackage{hyperref}
\usepackage{url}
\usepackage{booktabs}
\usepackage{amsfonts}
\usepackage{nicefrac}
\usepackage{microtype}
\usepackage{xcolor}

\usepackage{amsmath,amssymb}
\usepackage{graphicx}
\usepackage{caption}
\usepackage{subcaption}
\usepackage{float}
\usepackage{longtable}
\usepackage{array}
\usepackage{pdflscape}
\usepackage{enumitem}

\title{What Jobs Can AI Learn? \\
    Measuring Exposure by Reinforcement Learning}

\author{%
  Philip Moreira Tomei\thanks{Contact: philip@aiobjectives.org. We are grateful to Sam Manning, Pamela Mishkin, Julian Jacobs, Tom Cunningham, and Connacher Murphy for useful comments. Refine.ink was used to check the paper for consistency and clarity.} \\
  AI Objectives Institute \\
  \And
  Bouke Klein Teeselink \\
  AI Objectives Institute \\
  King's College London 
}

\begin{document}

\maketitle

\begin{abstract}
Which jobs can AI learn to do? We examine this for every occupation in the US economy. Existing indices measure the overlap between AI capabilities and occupational tasks rather than which tasks AI systems can learn to perform, and as a result misclassify occupations where the gap between present capability and learnability is large. Reinforcement learning in post-training, now the dominant paradigm at the frontier, is structured around task completion and maps more directly onto the task-based architecture of occupational classifications than prior approaches. Using LLM annotators guided by a rubric developed with RL experts, we evaluate 17,951 ONET tasks for training feasibility and aggregate to the occupation level, producing an RL Feasibility Index. The index diverges sharply from existing AI exposure measures for specific occupation groups: power plant operators, railroad conductors, and aircraft cargo handling supervisors score high on RL feasibility but low on general AI exposure, while musicians, chief executives, and microbiologists show the reverse. These divergences carry direct implications for policy interventions.
\end{abstract}

Measures of occupational exposure to AI have become central to labor market policy \citep{FreyOsborne2017TFSC, FeltenRajSeamans2018AEAPP, BrynjolfssonMitchellRock2018AEAPP, Webb2020SSRN, eloundou2024gpts}. The most widely cited, \citet{eloundou2024gpts}, finds that roughly 80\% of the US workforce has at least 10\% of their tasks exposed to large language models (LLMs), with the highest exposure among writers, analysts, and software developers according to one rubric.

Existing indices, however, might not sufficiently account for future improvements in AI capabilities. As such, policy makers who rely on these indices to tailor labor market policies and retraining programs may do so on incorrect or incomplete information. Hence, there is a strong need for a forward-looking measure that targets the source of improvements in AI capabilities to keep policymakers and economists up to date.

AI developers themselves have strong beliefs about where these improvements will come from. OpenAI reportedly expects the entire economy ``to become a Reinforcement Learning Machine", in which AI systems train on recordings of how professionals handle their day-to-day work \citep{palazzolo2025coworkers}. Meta has begun capturing employee mouse movements, keystrokes, and screenshots to train agents that perform work tasks autonomously \citep{paul2026meta}. Startups now build entire virtual workplaces, which are virtual machines outfitted with an email inbox, a Slack account, some coding tools and a web browser, in which firms can train an AI system to automate a white-collar job in its entirety \citep{roose2025mechanize}. No existing exposure measure captures which tasks can be turned into such training environments.

To fill this gap, we construct a new index based on reinforcement learning (RL), the training paradigm driving recent AI capability gains, covering every occupation in the U.S. economy. For the 17,951 tasks in the ONET database, LLM-based annotators first apply a binary physical feasibility gate (tasks requiring substantial physical embodiment receive a score of zero), then score RL training feasibility across eight dimensions, ranging from verification method to output tangibility, on a 1--10 scale, conditioned on occupation context. The average across these dimensions yields a task-level score; we then average across tasks within each occupation, weighting by ONET task-level importance ratings, to get an occupation-level score. The resulting index ranks every U.S. occupation by its exposure to RL-driven automation. Motivated by evidence that LLM-based occupational exposure scores can vary substantially across models \citep{YinVuPersico2026}, we assess robustness using two additional models and find very similar rankings. The index is publicly available at \url{https://github.com/boukektkcl/RL-exposure-public}.

Although our index strongly correlates with the exposure measures in \citet{eloundou2024gpts}, the two indices diverge for several occupations. Musicians, CEOs, and microbiologists rank high on LLM exposure but low on RL feasibility, because they tend to have subjective outputs,  environments that are hard to simulate, or need to perform physical tasks. Gas plant operators, power plant operators, and railroad conductors show the reverse (monitoring and control tasks with verifiable outcomes and simulable environments, but minimal text). The differences between the indices become much more pronounced when we only consider the subset of non-physical tasks. These divergences matter for policy, as workers in exposed occupations with low exposure scores for prior indices may fall outside current AI policy frameworks. Our index provides indications of where the next wave of automation pressure may concentrate, giving policymakers a forward-looking measure that other indices are likely to miss.

    We contribute to a rapidly expanding literature on AI exposure indices. A first generation measured exposure to automation or previous-generation AI \citep{FreyOsborne2017TFSC, ArntzGregoryZierahn2016OECD, NedelkoskaQuintini2018OECD, BrynjolfssonMitchellRock2018AEAPP, FeltenRajSeamans2021SMJ, Webb2020SSRN}. Since the introduction of ChatGPT in November 2022, the focus has shifted to generative AI. The index most-cited is \citet{eloundou2024gpts}, who score O*NET tasks on LLM exposure and create occupation-level LLM exposure measures based on those scores. \citet{ILO2023WP96} and \citet{Gmyrek2025} extend this approach globally and try to distinguish automation from augmentation. \citet{Pizzinelli2023IMF} adjust for task complementarity, finding that high-skill occupations are exposed but also strongly complemented. A parallel strand of research uses actual LLM usage data to measure  AI exposure: \citet{appel2025anthropic} analyze Claude conversations and \citet{tomlinson2025working} analyze Copilot interactions, both finding that AI use concentrates in information work. Our main contribution to this literature is to produce a forward looking index that considers which occupations are most likely to be exposed to further advances in AI capabilities.
    
  A closely related paper is \citet{FarboodiKohXia2026}, who develop a model of data-driven automation in which task-specific data accumulation determines the path of automation. They argue that verifiability and the density of reward signals are the main factors that influence how effectively tasks generate useful training data, which in turn improves the algorithm and expands the output of the task (and thus the training data). We complement their framework by constructing an index that measures the feasibility of RL-based training across tasks and aggregates these differences into occupational exposure to RL-driven automation.

Our work also relates to the task-based framework for analyzing technological change \citep{AutorLevyMurnane2003QJE, AcemogluAutor2011Handbook, AcemogluRestrepo2019JEP, acemoglu2022tasks} to a paradigm that has received no systematic occupational analysis. It also complements the growing empirical literature on AI and labor markets \citep{AcemogluAutorHazellRestrepo2022JLE,BrynjolfssonLiRaymond2025QJE,noy2023experimental,HuiReshefZhou2024OrgSci,klein2025generative,BrynjolfssonChandarChen2025,LichtingerHosseiniMaasoum2025,klein2026ai}. Where that work documents backward-looking evidence on early effects of AI adoption on employment and productivity, our index identifies the properties that make tasks amenable to the next wave of RL-driven automation.

\section{Methodology}
\label{sec:methodology}

We score RL feasibility at the task level, aggregate to occupations, and compare with existing AI exposure measures. We use O*NET 30.0, which contains 17,951 unique task statements across 894 occupations at the 8-digit SOC level. For each task, we ask how feasible it is to construct an RL environment in which AI could learn to perform it. The answer depends on properties such as whether the task admits a verifiable reward signal, a simulable environment, and a tractable decision space.

Before scoring any dimensions, we impose a binary physical feasibility gate that determines whether the task requires substantial physical interaction with the material world. Tasks that irreducibly require a physical body (manual labor, fine motor dexterity, locomotion) fail the gate and receive an RL index of 0, since we focus on RL environments that exist in software rather than physical AI. Tasks that can be performed primarily through digital means pass the gate and proceed to dimensional scoring.

For tasks that pass the gate, we decompose RL feasibility into eight dimensions, each scored on a 1--10 Likert scale (1 = RL infeasible, 10 = ideal for RL). The scoring assumes fine-tuning a pre-trained foundation model via RL methods such as reinforcement learning from human feedback (RLHF), reinforcement learning from AI feedback (RLAIF), or reinforcement learning with verifiable rewards (RLVR). For foundational treatments of these methods, see \citet{christiano2017deep,ouyang2022training,bai2022constitutional,rafailov2023direct}.

The dimensions are as follows. D1 (Verification Method Spectrum) captures where a task falls between deterministic verification (code that compiles) and contested professional judgment (psychotherapy outcomes), incorporating recent advances in rubric-based RL and AI-as-Judge. D2 (Environment Simulability) asks whether the task setting can be cheaply replicated digitally. The three MDP dimensions capture structural properties of the decision problem: whether task-relevant information is observable (D3), how varied task instances are and how broad the required expertise (D4), and how deep the sequential decision chain is (D5). D6 (Feedback Density \& Decomposability) measures both the timing and granularity of performance signals, reflecting advances in process reward models. D7 (Tool \& Interface Accessibility) captures whether the task can be performed through programmatic interfaces (APIs, CLIs) rather than GUIs. D8 (Output Tangibility) asks whether the task produces a concrete, inspectable artifact that can be graded independently of the process that created it. Full descriptions are in Appendix~\ref{app:prompt}.

The RL Feasibility Index for task $i$ is:
\begin{equation}
    \text{RL}_i = \frac{\bar{S}_i - 1}{9} \times 100, \quad \text{where } \bar{S}_i = \frac{1}{8}\sum_{d=1}^{8} S_{i,d}
    \label{eq:rl_index}
\end{equation}
and $S_{i,d} \in \{1,\ldots,10\}$ is the score on dimension $d$. The index maps all-ones to 0 and all-tens to 100. Tasks that fail the Physical Feasibility Gate receive $\text{RL}_i = 0$.

We submit all 17,951 occupation-task pairs for LLM-based evaluation. Each task is presented alongside its occupation title with the full rubric (Appendix~\ref{app:prompt}). The model first evaluates the physical feasibility gate, then performs a structured task reasoning step (classifying task type, identifying the core output, and predicting the binding constraint), and finally (if the task passes) returns eight scores with justifications.

The prompt is designed to be context-sensitive: it instructs annotators to condition scores on the occupation. ``Draft written correspondence'' receives different ratings for a Legal Secretary (routine, template-based) versus a Chief Executive (high-stakes strategic communications). Our primary annotator is Gemini 3.8 Flash, but we compare scores across Gemini 3.8 Flash, GPT-5.6 Luna and GLM-5.3 to assess the stability of our index with respect to model choice. All models are accessed via the OpenRouter API.\footnote{Gemini 3.8 Flash returns valid scores for 17,893 tasks; 58 tasks have no valid score. We exclude these tasks and renormalize importance weights within occupations. All 894 occupations retain at least one scored task. Appendix~\ref{app:model_comparison} reports coverage and comparisons across the three models.}

We aggregate from task-level exposure scores to occupations using O*NET task-importance weights:
\begin{equation}
    \overline{\text{RL}}_j = \sum_{i \in T_j} \frac{\text{Imp}_{i,j}}{\sum_{k \in T_j} \text{Imp}_{k,j}} \cdot \text{RL}_i
\end{equation}
where $T_j$ is the set of tasks in occupation $j$ with a valid score and $\text{Imp}_{i,j}$ is the O*NET importance rating (1--5 scale).

It is important to be clear about what our index is and is not measuring. We are specifically focusing on how feasible it is to improve an LLM's performance on a given task through RL-based post-training. As such, previous generations of AI such as rule-based software, predictive models, or classical machine learning fall outside our scope, as do tasks amenable to robotics or other forms of physical AI, as should be clear from our physical feasibility gate.

To interpret our index, note that labs create RL environments to train a system to perform a task on its own. Where such an environment is cheap to build and the nature of the task is amenable to reinforcement learning, the cost of a machine performing the task will fall, with outputs continuously improving in quality. Furthermore, with verifiable rewards, capabilities do not need to max out at the level of the best workers currently performing the task \citep{silverMasteringGameGo2017}. The index therefore provides an arguably more direct proxy for automation potential than many other AI exposure indices, which often combine automation and augmentation potential. The index does not, however, predict net employment in an occupation. That depends on the share of its tasks that cross the boundary, on whether the wage bill saved covers the cost of building the environment, and on how much cheaper output raises demand for the tasks that remain. Moreover, new or complementary tasks and jobs may emerge in adjacent areas to those that are exposed to RL. As such, our index is best interpreted as an indication of where displacement pressure from advances in RL may concentrate, rather than a forecast of net employment.

\section{Results}
\label{sec:results}

\subsection{Descriptives}
\label{sec:descriptives}

Table~\ref{tab:top_bottom_occ} reports the ten highest and lowest exposure occupations. The top of the distribution is dominated by clerical, data-processing, and information-handling roles (data entry keyers, proofreaders, payroll clerks, and order clerks) that operate in fully digital environments with rule-governed operations and verifiable outputs.

\begin{table}[!htbp]
\centering
\caption{Top and bottom 10 occupations by RL feasibility index. Model: Gemini 3.8 Flash}
\label{tab:top_bottom_occ}
\small
\begin{tabular}{p{9cm}r}
\toprule
\textbf{Occupation} & \textbf{Mean RL Index} \\
\midrule
\multicolumn{2}{l}{\textit{Panel A: Highest RL Feasibility}} \\
Data Entry Keyers & 76.08 \\
Proofreaders and Copy Markers & 75.51 \\
Payroll and Timekeeping Clerks & 75.16 \\
Order Clerks & 74.25 \\
Statistical Assistants & 72.03 \\
Travel Agents & 71.63 \\
Tax Preparers & 70.85 \\
Credit Authorizers, Checkers, and Clerks & 70.51 \\
Insurance Claims and Policy Processing Clerks & 70.18 \\
Data Warehousing Specialists & 70.17 \\
\midrule
\multicolumn{2}{l}{\textit{Panel B: Lowest RL Feasibility}} \\
Histology Technicians & 0.00 \\
Dining Room and Cafeteria Attendants and Bartender Helpers & 0.00 \\
Dishwashers & 0.00 \\
Maids and Housekeeping Cleaners & 0.00 \\
Agricultural Equipment Operators & 0.00 \\
Fallers & 0.00 \\
Stonemasons & 0.00 \\
Floor Sanders and Finishers & 0.00 \\
Cement Masons and Concrete Finishers & 0.00 \\
Terrazzo Workers and Finishers & 0.00 \\
\bottomrule
\end{tabular}
\caption*{\footnotesize \textit{Notes:} Importance-weighted means of task scores, 0--100. $N = 894$ occupations. Forty-seven occupations tie at zero; Panel B lists the first ten by O*NET-SOC code.}
\end{table}

The bottom of the distribution is dominated by occupations whose every task fails the physical feasibility gate. Dishwashers, stonemasons, floor sanders and finishers, and maids and housekeeping cleaners perform work that irreducibly requires a physical body; no RL environment can be constructed for tasks that have no digital representation. In general, 43.9\% of scored tasks fail the physical feasibility gate and receive a score of zero.

\begin{figure}[t!]
    \centering
    \caption{Mean RL Feasibility Index by SOC major group.}
    \label{fig:desc_soc}
    \includegraphics[width=0.8\textwidth]{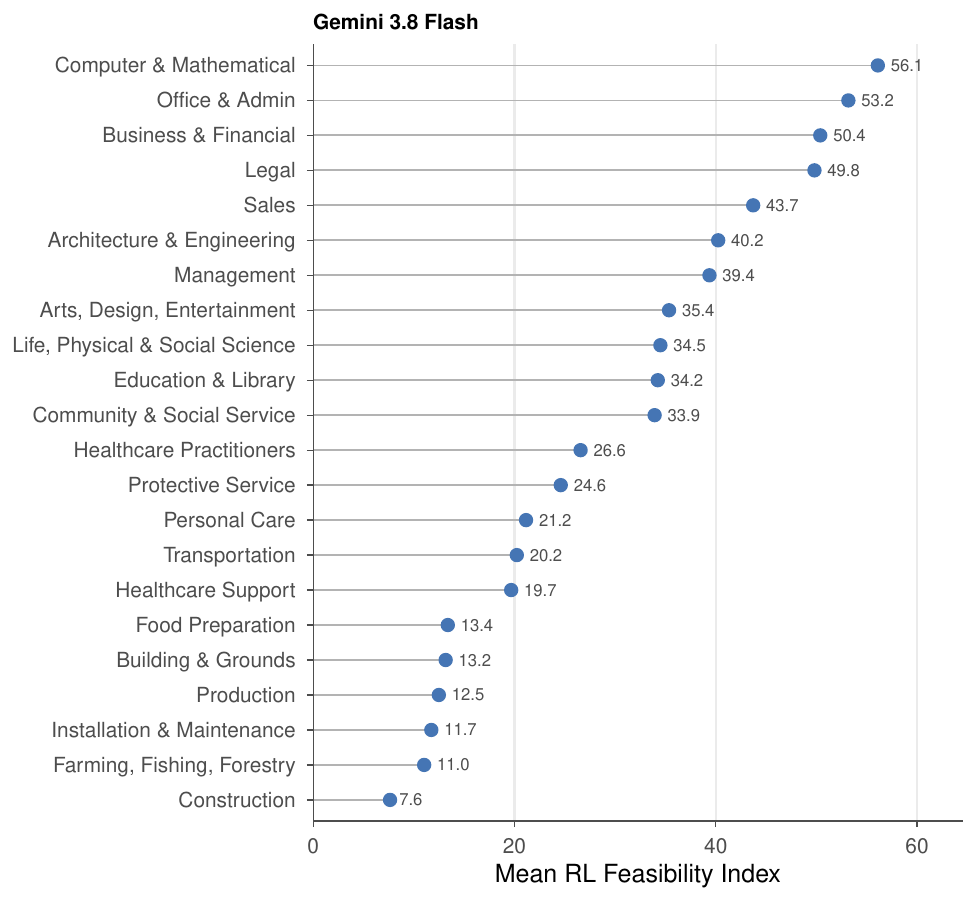}
    \caption*{\footnotesize \textit{Notes:} Each point is average occupation-level RL Feasibility Index score within the SOC major group. $N = 894$ occupations across 22 SOC major groups. Data from O*NET 30.0.}
\end{figure}

Figure~\ref{fig:desc_soc} plots mean RL feasibility by SOC major group. Computer and Mathematical occupations (SOC 15) have the highest mean (56.1), followed by Office and Administrative Support (53.2) and Business and Financial (50.4). Construction (7.6), Farming, Fishing, and Forestry (11.0), and Installation and Maintenance (11.7) score lowest, driven by high physical-gate failure rates.

Across the eight dimensions comprising the index, output tangibility and tool accessibility score highest, whereas task variability and verification score lowest. Correlations are generally positive, although tool accessibility is weakly related to the other dimensions and has a small negative correlation with task variability. A principal component analysis finds that the first principal component explains 69.0\% of the variation, and loads positively on all dimensions, while PC2 explains 13.2\% and is dominated by tool accessibility. Appendix~\ref{app:descriptives} contains more detailed descriptives.

\subsection{Labor Market Profiles}
\label{sec:labour_market}

To gain further insights into the job characteristics of high vs. low exposure occupations, we link our occupation-level index to individual-level labor market data from Revelio Labs. Revelio Labs is a provider of workforce analytics derived from public professional profiles, and the US dataset contains position records with information on occupation (O*NET-SOC code), salary, seniority level (1 = entry-level through 7 = executive), and employer. We aggregate the Revelio data to O*NET occupation codes, and merge it with our RL index. The main variables we consider are wages, seniority, and industry.

We report results for two samples: positions active on 1 November 2025 and positions active on 1 October 2022 (right before the release of ChatGPT).  While the former has the advantage of being more recent, there is also a risk that RL exposure already had labor market effects, which would mean that characteristics such as wages and seniority may be influenced by RL exposure. We therefore report both. The figures below show the recent sample, but the pre-ChatGPT results are nearly identical and are reported in Appendix~\ref{app:pre_chatgpt}.

Figure~\ref{fig:wage} shows RL feasibility across the wage distribution. Exposure rises from 34.2 in the lowest decile to a peak of 42.7 in the eighth decile, then declines to 39.0 in the highest decile. Exposure at the top is thus higher than exposure at the bottom. This pattern is consistent with some prior waves of automation, which predominantly eroded middle-income positions, leading to polarization of the wage distribution and rising inequality \citep{AutorKatzKearney2006QJE, AutorDorn2013AER, GoosManning2007REStat}.

\begin{figure}[t!]
    \centering
    \caption{Mean RL Feasibility Index by wage decile (November 2025).}
    \label{fig:wage}
    \includegraphics[width=0.8\textwidth]{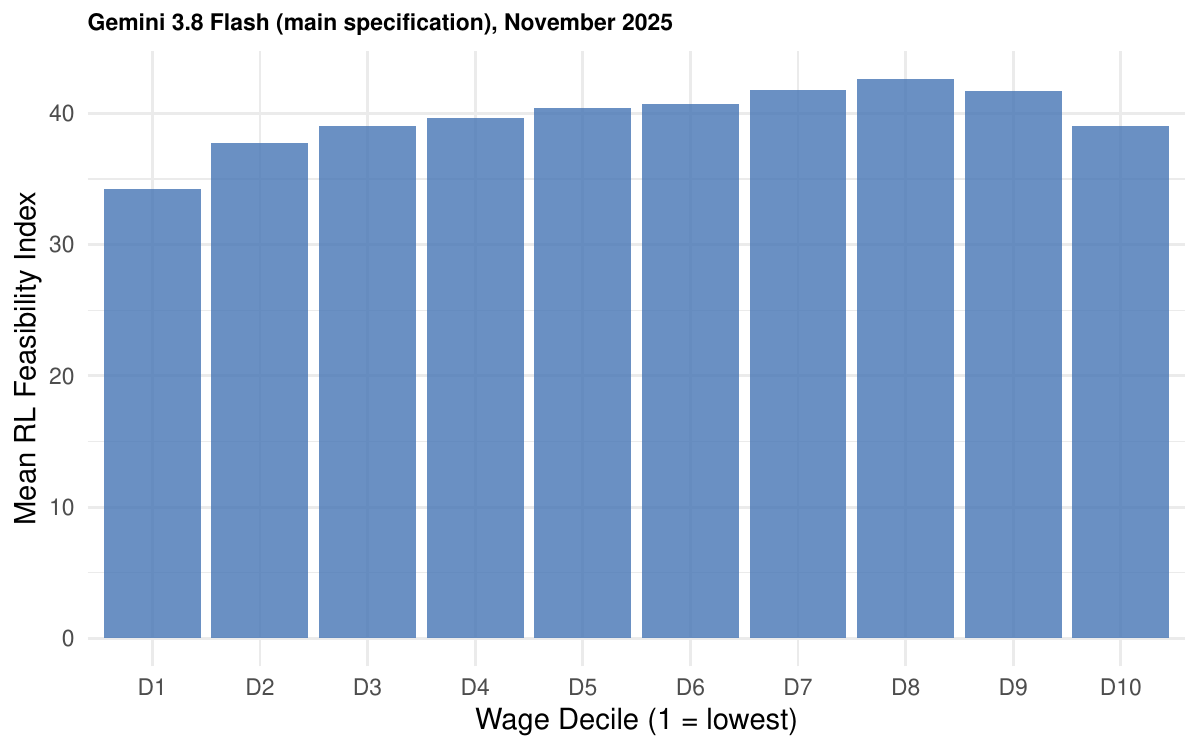}
    \caption*{\footnotesize \textit{Notes:} Bars show the mean RL Feasibility Index by wage decile. Decile 1 is lowest-paid. Wage deciles are constructed from individual position salaries using Revelio Labs position records active on November 2025 (94.5 million records), merged with our occupation-level index by O*NET-SOC code.}
\end{figure}

\begin{figure}[t!]
    \centering
    \caption{Mean RL Feasibility Index by seniority level (November 2025).}
    \label{fig:seniority}
    \includegraphics[width=0.8\textwidth]{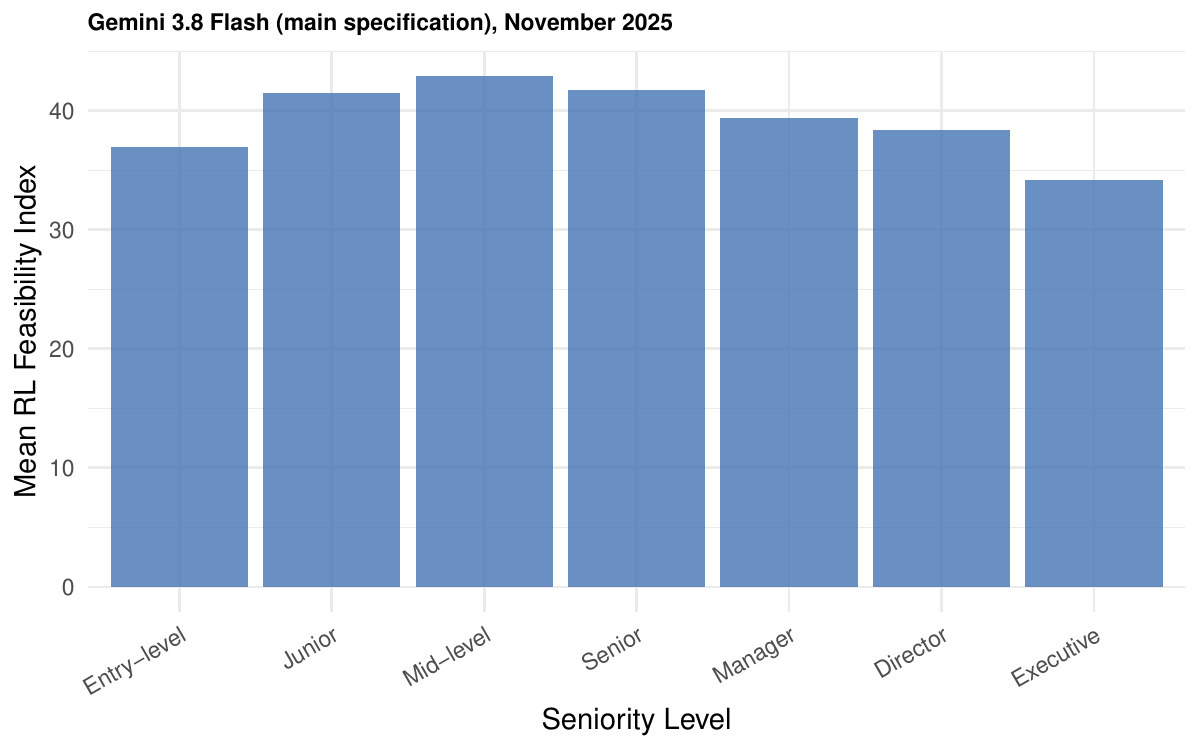}
    \caption*{\footnotesize \textit{Notes:} Bars show the employment-weighted mean RL Feasibility Index by seniority level. Seniority levels are from Revelio Labs position records active in November 2025 (94.5 million records), linked to our occupation-level index by O*NET-SOC code. Each position receives its occupation's exposure score. The matched sample covers 888 occupations.}
\end{figure}

Figure~\ref{fig:seniority} shows RL exposure by seniority level. Similar to our wage results, RL feasibility shows an inverted U-shaped relationship with seniority. Exposure rises from entry-level (37.0) through mid-level positions (42.9), then declines to its minimum at the executive level (34.2). Indeed, the most junior and most senior workers are least exposed, while mid-career workers face the highest RL feasibility. 

Table~\ref{tab:regression_recent} reports occupation-level regressions of RL feasibility on log mean salary and a quadratic in mean seniority.\footnote{Seniority levels vary within occupations. We average the seniority level (coded on a discrete 7-point scale) across the positions in each occupation. Table~\ref{tab:regression_recent} uses positions active on 1 November 2025 and Appendix Table~\ref{tab:regression_pre_chatgpt} uses positions active on 1 October 2022.} The salary coefficient is not statistically distinguishable from zero: it is $-2.4$ in the OLS specification and $2.6$ with SOC major-group fixed effects. Consistent with the figures above, the seniority terms imply an inverted-U relationship, with linear coefficients of 35.0 and 20.5 and quadratic coefficients of $-4.8$ and $-3.4$, respectively (all $p < 0.01$). The fitted peaks occur at mean seniority levels of 3.6 and 3.0. Thus, the seniority pattern persists within SOC major groups, whereas we find no clear independent association with mean salary in these specifications. Appendix Table~\ref{tab:regression_pre_chatgpt} reports similar results for the pre-ChatGPT sample.

\begin{table}[!htbp] \centering 
  \caption{Occupation-level regressions of RL feasibility on wage and seniority (November 2025). Model: Gemini 3.8 Flash} 
  \label{tab:regression_recent} 
\scriptsize 
\begin{tabular}{@{\extracolsep{5pt}}lcc} 
\\[-1.8ex]\hline 
\hline \\[-1.8ex] 
\\[-1.8ex] & \multicolumn{2}{c}{RL Feasibility Index} \\ 
 & OLS & SOC major FE \\ 
\hline \\[-1.8ex] 
 Log(mean salary) & $-$2.431 & 2.603 \\ 
  & (2.701) & (2.332) \\ 
  & & \\ 
 Mean seniority & 35.043$^{***}$ & 20.504$^{***}$ \\ 
  & (4.048) & (2.922) \\ 
  & & \\ 
 Mean seniority squared & $-$4.844$^{***}$ & $-$3.426$^{***}$ \\ 
  & (0.704) & (0.502) \\ 
  & & \\ 
 Constant & 6.367 &  \\ 
  & (27.017) &  \\ 
  & & \\ 
\hline \\[-1.8ex] 
SOC major group FE & No & Yes \\ 
Observations & 888 & 888 \\ 
R$^{2}$ & 0.230 & 0.668 \\ 
Adjusted R$^{2}$ & 0.227 & 0.658 \\ 
\hline 
\hline \\[-1.8ex] 
\textit{Note:}  & \multicolumn{2}{r}{$^{*}$p$<$0.1; $^{**}$p$<$0.05; $^{***}$p$<$0.01} \\ 
\end{tabular} 
\caption*{\footnotesize \textit{Notes:} OLS and SOC-major-group fixed-effects regressions of the RL feasibility index of model Gemini 3.8 Flash on log mean salary and a quadratic in mean seniority. Unit of observation is an O*NET occupation ($N = 888$). Salary and seniority are computed from Revelio Labs position records active in November 2025. Standard errors in parentheses. $^{*}p<0.1$; $^{**}p<0.05$; $^{***}p<0.01$.}
\end{table}

\subsection{Comparison across Annotation Models}
\label{sec:model_comparison}

	LLM-based exposure scores sometimes vary heavily with the model. \citet{YinVuPersico2026} apply the \citet{eloundou2024gpts} rubric to the O*NET 30.0 tasks with GPT-5, Gemini 2.5, and Claude 4.5 and find that the models agree on only 57 to 73\% of task labels, with mean exposure ranging from 0.14 to 0.51.	We therefore assess whether the index depends on the choice of annotator by scoring the same O*NET tasks with GPT-5.6 Luna and GLM-5.3. 
	
	The results show that the occupation-level indices are strongly correlated across models. The correlation between our main Gemini 3.8 Flash model and GPT-5.6 Luna is 0.971, and with GLM-5.3 it is 0.965. All three models assign the highest mean exposure to Computer and Mathematical occupations, followed by Office and Administrative Support, and the lowest to Construction. However, GLM-5.3 assigns higher scores overall, with an occupation-level mean of 33.1 compared with 29.0 for Gemini 3.8 Flash and 29.3 for GPT-5.6 Luna.

Consistency between LLM annotators is similarly high at the task level, with correlations ranging from 0.868 to 0.898. Moreover, physical-gate decisions agree for 93.6--95.3\% of tasks. Taken together, we conclude that our rubric and index are highly robust to the exact model used for scoring.

\subsection{Comparison to Eloundou et al.}
\label{sec:comparison}

Next, to identify where our index differs from existing indices, we compare our index with four exposure measures from \citet{eloundou2024gpts}. Their annotators ask whether an LLM can reduce the time to complete each O*NET task by at least 50\% while maintaining quality. A task is labeled E0 if no LLM can, E1 if an LLM on its own can, and E2 if an LLM can only with additional software built on top of it. Their $\alpha$ measure only considers E1 tasks ($E1=1$ and $E2=0$), their $\beta$ measure adds E2 tasks at half the weight ($E1=1$ and $E2=0.5$), and their $\gamma$ measure adds them at full weight ($E1=1$ and $E2=1$). We use the LLM-annotated versions of $\alpha$, $\beta$, and $\gamma$, and add their human-annotated version of $\beta$ as an additional comparison measure. We aggregate all four measures to occupations with the same O*NET importance weights as our index. For each comparison, we calculate the correlation between our index and their indices, both for the full set of tasks and for the subset of tasks that pass the physical gate. We then examine for which occupations our index diverges from theirs. 

Table~\ref{tab:eloundou_alternatives} in Appendix~\ref{app:eloundou_alternatives} reports the occupation-level correlations. Aggregating over all tasks, the Pearson correlation between our RL feasibility index and the exposure measures of \citet{eloundou2024gpts} is 0.576 for $\alpha$, 0.912 for $\beta$, 0.888 for $\gamma$, and 0.854 for human-annotated $\beta$. Rank correlations are similar. Our index is thus more closely correlated with measures that include capabilities based on complementary software ($\beta$ and $\gamma$) than with exposure to a standalone LLM ($\alpha$). The correlation with human-annotated $\beta$ is also high, showing that the relationship holds when the benchmark is based on human judgments. However, the correlations are lower when we restrict both indices to the 10,011 matched tasks that pass the physical feasibility gate. The Pearson correlations are 0.549 for $\alpha$, 0.379 for $\beta$, and drop to 0.005 for $\gamma$, and $-0.102$ for human-annotated $\beta$, with rank correlations showing the same ordering. This suggests that a significant part of the full-sample correlation is driven by agreement on low exposure of physical tasks across the measures. Among tasks that pass the gate, exposure to LLMs and feasibility for RL training appear to capture different features of work.

\begin{figure}[t!]
    \centering
    \caption{Occupation-level RL feasibility vs.\ general AI exposure ($\beta$). Solid lines mark medians; shaded regions denote the four quadrants. Selected occupations are labeled in each quadrant.}
    \label{fig:quadrant}
    \includegraphics[width=0.8\textwidth]{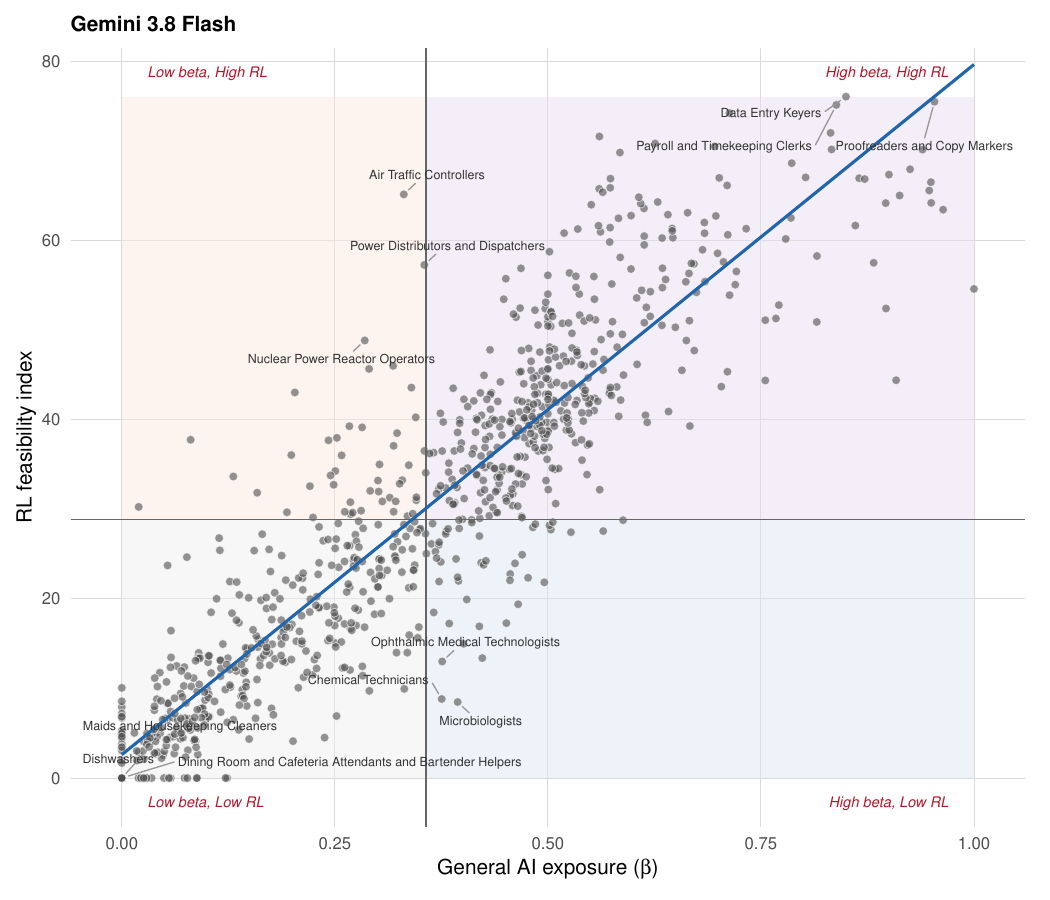}
    \caption*{\footnotesize \textit{Notes:} Each point is one occupation ($N = 894$). The horizontal axis is the importance-weighted mean of \citet{eloundou2024gpts} $\beta$ scores (0 = not exposed, 0.5 = exposed with software, 1 = exposed to standalone LLMs). The vertical axis is our RL Feasibility Index (0--100). Solid lines mark sample medians; shaded regions denote the four quadrants defined by above/below-median splits on each axis. Selected occupations furthest from the medians in each quadrant are labeled. Both indices are aggregated to occupations using O*NET 30.0 importance weights.}
\end{figure}

Figure~\ref{fig:quadrant} plots RL feasibility against $\beta$ for each occupation based on the full set of tasks. Appendix Figures~\ref{fig:quadrant_alpha} to \ref{fig:quadrant_beta_human} show the same plot for $\alpha$, $\gamma$, and human-annotated $\beta$, and Appendix Figures~\ref{fig:quadrant_beta_gated} to \ref{fig:quadrant_beta_human_gated} show the corresponding plots for the subset of tasks that pass the physical gate. We divide occupations into four quadrants of above versus below median exposure on our index, and above vs. below median exposure on Eloundou et al.'s $\beta$. 

While the correlation between the measures is high when we consider the full task list, we nonetheless observe interesting deviations. The lower-right quadrant of Figure~\ref{fig:quadrant} contains occupations that score high on general AI exposure but low on RL feasibility. Half of them are life, physical, and social science occupations or healthcare practitioners. Microbiologists, chemical technicians, biochemists, medical scientists, and hospitalists combine physical and/or patient work, which fails the physical feasibility gate, with documentation and analysis, which are exposed to LLMs. The quadrant also contains engineering, arts, and leadership roles, such as nanosystems engineers, musicians, curators, and chief executives, whose digital outputs have subjective success criteria and environments that resist simulation. Thus, the divergence reflects both difficulties in training digitally mediated tasks and differences in how the indices treat embodied work. These conclusions are similar for $\alpha$, $\gamma$, and human-annotated $\beta$ (see Appendix Figures~\ref{fig:quadrant_alpha}, \ref{fig:quadrant_gamma}, and \ref{fig:quadrant_beta_human}, respectively).

The upper-left quadrant shows the reverse, occupations with low general AI exposure but high RL feasibility. The largest group consists of monitoring and control roles: air traffic controllers, power distributors and dispatchers, nuclear power reactor operators, power plant operators, gas plant operators, railroad conductors and yardmasters, and aircraft cargo handling supervisors. Their work is not text-centric, yet it has the structural features that RL exploits, namely verifiable outcomes, discrete action spaces, shallow decision chains, and immediate feedback from instrumented systems. Railroad conductors, power plant operators, gas plant operators, and biomass power plant managers fall in this quadrant under all four measures, and air traffic controllers and aircraft cargo handling supervisors under three of the four. A second group consists of counseling occupations (mental health counselors, marriage and family therapists, and substance abuse counselors), which also fall in this quadrant under all four measures but with RL scores only 5 to 7 points above the median. From a policy perspective, the monitoring and control group is arguably the most relevant, as it receives low exposure scores under existing AI exposure frameworks but may nonetheless face clear automation risk.

Appendix Table~\ref{tab:quadrant_descriptives} gives additional details about the workers in each quadrant based on the same set of Revelio Labs LinkedIn positions on 1 October 2022 from Section~\ref{sec:labour_market}. The high-$\beta$-low-RL quadrant comprises 5.4\% of the 93.7 million positions and the low-$\beta$-high-RL quadrant 3.3\%. While this is a relatively small proportion of the total, it still amounts to 8.2 million positions, which is an economically significant part of the economy that is missed by existing exposure indices.\footnote{The fraction of off-diagonal jobs increases significantly if we only consider tasks that pass the physical gate.} Workers in the lower-right quadrant earn an average salary of \$122,970, against \$93,783 across all positions, and 44.2\% of them hold manager-level or higher positions, against 18.4\% overall.\footnote{Chief executives, who account for 45.8\% of the positions in this quadrant, are the main driver of these numbers.} Workers in the upper-left quadrant on average earn a salary of \$75,798 and 11.5\% of them hold manager-level or higher positions. Occupations with high RL feasibility but low general exposure thus employ workers who earn less, hold more junior positions, and work outside large cities more often than the average worker in our sample. This suggests that these workers, who are most exposed to RL but fall outside traditional exposure measures, are likely also the least adaptable ones \citep{manning2025adaptable}.

Appendix Figures~\ref{fig:quadrant_beta_gated} to \ref{fig:quadrant_beta_human_gated} show what this disagreement looks like for gate-passing tasks. The number of off-diagonal occupations increases, and the low-exposure, high-RL quadrant  now mostly consists of physical occupations that retain only a few digital tasks that can easily be turned into RL environments. Among occupations that keep at least half of their tasks, the same monitoring and control roles as in the full sample remain in this quadrant. The high-exposure, low-RL quadrant now consists of occupations whose work is almost entirely digital, with a median of 16 gate-passing tasks, and it is dominated by research, engineering, and writing occupations. Biochemists, physicists, medical scientists, editors, and writers and authors fall in this quadrant. In other words, the indices mostly disagree about knowledge and creative work, where an LLM saves time but the output is hard to verify or to simulate.

\section{Conclusion}
\label{sec:discussion}
Most existing AI exposure indices ask what large language models can currently do. We construct a new index that measures how exposed each occupation is to automation through reinforcement learning, the training paradigm behind recent AI capability gains. The RL exposure of a job depends on task-level properties such as the degree to which the task has verifiable rewards, a simulable environment, and tractable decision spaces. We create this index for all US occupations. RL feasibility peaks in the eighth wage decile and at mid-level seniority, declining at higher wages and seniority levels. The highest wage decile remains more exposed than the lowest, while executives have the lowest exposure across seniority levels.

We compare our index with \citet{eloundou2024gpts}. Across occupations, our Gemini 3.8 Flash index correlates with Eloundou et al.'s $\beta$ measure at 0.912; the correlation is 0.379 among occupations built from tasks that pass the physical-feasibility screen. Occupations that score high on general AI exposure but low on RL feasibility include CEOs and musicians, whose work can involve subjective success criteria and environments that resist simulation, as well as microbiologists, whose work includes embodied laboratory tasks. The reverse group (low general AI exposure but high RL feasibility) consists of monitoring and control occupations (gas plant operators, railroad conductors, aircraft cargo supervisors) whose tasks are not text-centric but have features that RL exploits: verifiable outcomes, discrete action spaces, and immediate feedback from instrumented systems. From a policy perspective, this second group may be the most consequential, as these workers fall outside existing AI exposure frameworks yet face clear automation risk.

The concentration of RL feasibility among upper-middle-wage occupations echoes earlier research on routine-biased technological change, which shows that previous automation often eroded the middle class, leading to increased inequality \citep{AutorKatzKearney2006QJE, GoosManning2007REStat, AutorDorn2013AER}. Combined with LLM-driven displacement of knowledge work at the top of the distribution \citep{klein2025generative}, RL exposure in these occupations suggests a broader potential reach of automation.

It is important to note that we measure \textit{exposure} to automation from reinforcement learning, which is likely to differ from actual automation. Whether a high-scoring task is actually automated depends on costs, wages, and adoption frictions such as legal constraints. Building an RL environment is a fixed cost, so a task is automated only when it is performed at enough scale to amortize that cost \citep{LashkariLiQiuThompson2026}. Some high-scoring occupations are also among the lower-paid, which weakens the business case for investment. Our index identifies \textit{where} RL can be used, whereas labor costs and capital investment determine \textit{when}.

Our index has limitations. It relies on LLM-generated annotations, which may be biased \citep{zheng2023judging}, and these biases may correlate with Eloundou et al.'s LLM-based scores. The correlation with their human-annotated $\beta$ measure is high across all occupations but slightly negative in the gate-passing sample, so the comparison does not establish agreement within the digitally feasible domain. In addition, we cover only the U.S.\ occupational structure. Validation against observed automation adoption and extensions to other labor markets are natural next steps.

\section*{Data Availability}
The task-level RL Feasibility Index data is publicly available at \url{https://github.com/boukektkcl/RL-exposure-public}. Revelio Labs data are proprietary and cannot be redistributed.\\

\bibliographystyle{plainnat}
\bibliography{Bibliography}

@article{acemoglu2022tasks,
	title={Tasks, Automation, and the Rise in {U.S.} Wage Inequality},
	author={Acemoglu, Daron and Restrepo, Pascual},
	journal={Econometrica},
	volume={90},
	number={5},
	pages={1973--2016},
	year={2022},
	doi={10.3982/ECTA19815}
}

@article{AcemogluAutorHazellRestrepo2022JLE,
	author    = {Acemoglu, Daron and Autor, David and Hazell, Jonathon and Restrepo, Pascual},
	title     = {{Artificial Intelligence} and Jobs: Evidence from Online Vacancies},
	journal   = {Journal of Labor Economics},
	year      = {2022},
	volume    = {40},
	number    = {S1},
	pages     = {S293--S340},
	note      = {Special Issue},
	doi       = {10.1086/718327}
}

@article{AcemogluRestrepo2019JEP,
	author    = {Acemoglu, Daron and Restrepo, Pascual},
	title     = {Automation and New Tasks: How Technology Displaces and Reinstates Labor},
	journal   = {Journal of Economic Perspectives},
	year      = {2019},
	volume    = {33},
	number    = {2},
	pages     = {3--30},
	doi       = {10.1257/jep.33.2.3}
}

@misc{appel2025anthropic,
	title={{Anthropic} Economic Index Report: Uneven Geographic and Enterprise {AI} Adoption},
	author={Appel, Ruth and McCrory, Peter and Tamkin, Alex and McCain, Miles and Neylon, Tyler and Stern, Michael},
	year={2025},
	howpublished={arXiv preprint arXiv:2511.15080},
	eprint={2511.15080},
	archiveprefix={arXiv}
}

@techreport{ArntzGregoryZierahn2016OECD,
	author      = {Arntz, Melanie and Gregory, Terry and Zierahn, Ulrich},
	title       = {The Risk of Automation for Jobs in {OECD} Countries: A Comparative Analysis},
	institution = {OECD Publishing},
	type        = {OECD Social, Employment and Migration Working Papers},
	number      = {189},
	year        = {2016},
	doi         = {10.1787/5jlz9h56dvq7-en}
}

@article{AutorKatzKearney2006QJE,
	title={The Polarization of the {U.S.} Labor Market},
	author={Autor, David H. and Katz, Lawrence F. and Kearney, Melissa S.},
	journal={American Economic Review},
	volume={96},
	number={2},
	pages={189--194},
	year={2006},
	doi={10.1257/000282806777212620}
}

@article{AutorLevyMurnane2003QJE,
	author    = {Autor, David H. and Levy, Frank and Murnane, Richard J.},
	title     = {The Skill Content of Recent Technological Change: An Empirical Exploration},
	journal   = {The Quarterly Journal of Economics},
	year      = {2003},
	volume    = {118},
	number    = {4},
	pages     = {1279--1333},
	doi       = {10.1162/003355303322552801}
}

@misc{BrynjolfssonChandarChen2025,
	author       = {Brynjolfsson, Erik and Chandar, Bharat and Chen, Ruyu},
	title        = {Canaries in the Coal Mine? {Six} Facts about the Recent Employment Effects of {Artificial Intelligence}},
	howpublished = {Stanford Digital Economy Lab Working Paper},
	year         = {2025},
	note         = {Revised August 2026},
	url          = {https://digitaleconomy.stanford.edu/publications/canaries-in-the-coal-mine/}
}

@article{BrynjolfssonLiRaymond2025QJE,
	author    = {Brynjolfsson, Erik and Li, Danielle and Raymond, Lindsey},
	title     = {Generative {AI} at Work},
	journal   = {The Quarterly Journal of Economics},
	year      = {2025},
	volume    = {140},
	number    = {2},
	pages     = {889--942},
	doi       = {10.1093/qje/qjae044}
}

@article{BrynjolfssonMitchellRock2018AEAPP,
	author    = {Brynjolfsson, Erik and Mitchell, Tom and Rock, Daniel},
	title     = {What Can Machines Learn, and What Does It Mean for Occupations and the Economy?},
	journal   = {AEA Papers and Proceedings},
	year      = {2018},
	volume    = {108},
	pages     = {43--47},
	doi       = {10.1257/pandp.20181019}
}

@article{eloundou2024gpts,
	title={{GPTs} are {GPTs}: Labor Market Impact Potential of {LLMs}},
	author={Eloundou, Tyna and Manning, Sam and Mishkin, Pamela and Rock, Daniel},
	journal={Science},
	volume={384},
	number={6702},
	pages={1306--1308},
	year={2024},
	doi={10.1126/science.adj0998}
}

@techreport{FarboodiKohXia2026,
	author    = {Farboodi, Maryam and Koh, Andrew J. and Xia, Anchi},
	title     = {Data-Driven Automation},
	year      = {2026},
	month     = jun,
	institution = {National Bureau of Economic Research},
	type      = {Working Paper},
	number    = {35320},
	doi       = {10.3386/w35320},
	url       = {https://www.nber.org/papers/w35320}
}

@article{FeltenRajSeamans2018AEAPP,
	author    = {Felten, Edward W. and Raj, Manav and Seamans, Robert},
	title     = {A Method to Link Advances in {Artificial Intelligence} to Occupational Abilities},
	journal   = {AEA Papers and Proceedings},
	year      = {2018},
	volume    = {108},
	pages     = {54--57},
	doi       = {10.1257/pandp.20181021}
}

@article{FeltenRajSeamans2021SMJ,
	author    = {Felten, Edward and Raj, Manav and Seamans, Robert},
	title     = {Occupational, Industry, and Geographic Exposure to {Artificial Intelligence}: A Novel Dataset and Its Potential Uses},
	journal   = {Strategic Management Journal},
	year      = {2021},
	volume    = {42},
	number    = {12},
	pages     = {2195--2217},
	doi       = {10.1002/smj.3286}
}

@article{FreyOsborne2017TFSC,
	author    = {Frey, Carl Benedikt and Osborne, Michael A.},
	title     = {The Future of Employment: How Susceptible Are Jobs to Computerisation?},
	journal   = {Technological Forecasting and Social Change},
	year      = {2017},
	volume    = {114},
	pages     = {254--280},
	doi       = {10.1016/j.techfore.2016.08.019}
}

@techreport{Gmyrek2025,
	author       = {Gmyrek, Pawe{\l} and Berg, Janine and Kami{\'n}ski, Karol and Konopczy{\'n}ski, Filip and {\L}adna, Agnieszka and Nafradi, Balint and Ros{\l}aniec, Konrad and Troszy{\'n}ski, Marek},
	title        = {Generative {AI} and Jobs: A Refined Global Index of Occupational Exposure},
	institution  = {International Labour Organization},
	type         = {ILO Working Paper},
	number       = {140},
	address      = {Geneva},
	year         = {2025},
	month        = may,
	doi          = {10.54394/HETP0387},
	isbn         = {978-92-2-042185-7}
}

@article{AutorDorn2013AER,
	author  = {Autor, David H. and Dorn, David},
	title   = {The Growth of Low-Skill Service Jobs and the Polarization of the {US} Labor Market},
	journal = {American Economic Review},
	year    = {2013},
	volume  = {103},
	number  = {5},
	pages   = {1553--1597},
	doi     = {10.1257/aer.103.5.1553},
}

@article{GoosManning2007REStat,
	author  = {Goos, Maarten and Manning, Alan},
	title   = {Lousy and Lovely Jobs: The Rising Polarization of Work in {Britain}},
	journal = {The Review of Economics and Statistics},
	year    = {2007},
	volume  = {89},
	number  = {1},
	pages   = {118--133},
	doi     = {10.1162/rest.89.1.118},
}

@article{HuiReshefZhou2024OrgSci,
	author  = {Hui, Xiang and Reshef, Oren and Zhou, Luofeng},
	title   = {The Short-Term Effects of Generative {Artificial Intelligence} on Employment: Evidence from an Online Labor Market},
	journal = {Organization Science},
	volume  = {35},
	number  = {6},
	pages   = {1977--1989},
	year    = {2024},
	doi     = {10.1287/orsc.2023.18441}
}

@techreport{ILO2023WP96,
	author      = {Gmyrek, Pawe{\l} and Berg, Janine and Bescond, David},
	title       = {Generative {AI} and Jobs: A Global Analysis of Potential Effects on Job Quantity and Quality},
	institution = {International Labour Organization},
	type        = {ILO Working Paper},
	number      = {96},
	address     = {Geneva},
	year        = {2023},
	month       = aug,
	doi         = {10.54394/FHEM8239}
}

@inproceedings{christiano2017deep,
  author    = {Christiano, Paul F. and Leike, Jan and Brown, Tom B. and Martic, Miljan and Legg, Shane and Amodei, Dario},
  title     = {Deep Reinforcement Learning from Human Preferences},
  booktitle = {Advances in Neural Information Processing Systems},
  volume    = {30},
  year      = {2017}
}

@inproceedings{ouyang2022training,
	author    = {Ouyang, Long and Wu, Jeff and Jiang, Xu and Almeida, Diogo and Wainwright, Carroll L. and Mishkin, Pamela and Zhang, Chong and Agarwal, Sandhini and Slama, Katarina and Ray, Alex and Schulman, John and Hilton, Jacob and Kelton, Fraser and Miller, Luke and Simens, Maddie and Askell, Amanda and Welinder, Peter and Christiano, Paul F. and Leike, Jan and Lowe, Ryan},
	title     = {Training Language Models to Follow Instructions with Human Feedback},
	booktitle = {Advances in Neural Information Processing Systems},
	volume    = {35},
	pages     = {27730--27744},
	year      = {2022},
	doi       = {10.52202/068431-2011}
}

@techreport{YinVuPersico2026,
	author      = {Yin, Michelle and Vu, Hoa and Persico, Claudia},
	title       = {How (un)Stable Are {LLM} Occupational Exposure Scores? {E}vidence from Multi-Model Replication},
	institution = {National Bureau of Economic Research},
	type        = {Working Paper},
	number      = {35110},
	year        = {2026},
	month       = apr,
	doi         = {10.3386/w35110}
}

@inproceedings{zheng2023judging,
	author    = {Zheng, Lianmin and Chiang, Wei-Lin and Sheng, Ying and Zhuang, Siyuan and Wu, Zhanghao and Zhuang, Yonghao and Lin, Zi and Li, Zhuohan and Li, Dacheng and Xing, Eric P. and Zhang, Hao and Gonzalez, Joseph E. and Stoica, Ion},
	title     = {Judging {LLM}-as-a-Judge with {MT-Bench} and {Chatbot Arena}},
	booktitle = {Advances in Neural Information Processing Systems},
	volume    = {36},
	pages     = {46595--46623},
	year      = {2023},
	doi       = {10.52202/075280-2020}
}

@misc{bai2022constitutional,
  author    = {Bai, Yuntao and Kadavath, Saurav and Kundu, Sandipan and Askell, Amanda and Kernion, Jackson and Jones, Andy and Chen, Anna and Goldie, Anna and Mirhoseini, Azalia and McKinnon, Cameron and Chen, Carol and Olsson, Catherine and Olah, Christopher and Hernandez, Danny and Drain, Dawn and Ganguli, Deep and Li, Dustin and Tran-Johnson, Eli and Perez, Ethan and Kerr, Jamie and Mueller, Jared and Ladish, Jeffrey and Landau, Joshua and Ndousse, Kamal and Lukosuite, Kamile and Lovitt, Liane and Sellitto, Michael and Elhage, Nelson and Schiefer, Nicholas and Mercado, Noemi and DasSarma, Nova and Lasenby, Robert and Larson, Robin and Ringer, Sam and Johnston, Scott and Kravec, Shauna and {El Showk}, Sheer and Fort, Stanislav and Lanham, Tamera and Telleen-Lawton, Timothy and Conerly, Tom and Henighan, Tom and Hume, Tristan and Bowman, Samuel R. and Hatfield-Dodds, Zac and Mann, Ben and Amodei, Dario and Joseph, Nicholas and McCandlish, Sam and Brown, Tom and Kaplan, Jared},
  title     = {Constitutional {AI}: Harmlessness from {AI} Feedback},
  year      = {2022},
  howpublished={arXiv preprint arXiv:2212.08073},
  eprint    = {2212.08073},
  archiveprefix = {arXiv}
}

@inproceedings{rafailov2023direct,
	author    = {Rafailov, Rafael and Sharma, Archit and Mitchell, Eric and Manning, Christopher D. and Ermon, Stefano and Finn, Chelsea},
	title     = {Direct Preference Optimization: Your Language Model is Secretly a Reward Model},
	booktitle = {Advances in Neural Information Processing Systems},
	volume    = {36},
	pages     = {53728--53741},
	year      = {2023},
	doi       = {10.52202/075280-2338}
}

@techreport{klein2026ai,
	author      = {{Klein Teeselink}, Bouke and Carey, Daniel},
	title       = {{AI}, Automation, and Expertise},
	institution = {SSRN},
	type        = {Working Paper},
	number      = {6134506},
	year        = {2026},
	month       = jan,
	doi         = {10.2139/ssrn.6134506}
}

@techreport{klein2025generative,
	author      = {{Klein Teeselink}, Bouke},
	title       = {Generative {AI} and Labor Market Outcomes: Evidence from the {United Kingdom}},
	institution = {SSRN},
	type        = {Working Paper},
	number      = {5516798},
	year        = {2025},
	month       = sep,
	doi         = {10.2139/ssrn.5516798}
}

@techreport{LichtingerHosseiniMaasoum2025,
	author      = {{Hosseini Maasoum}, Seyed Mahdi and Lichtinger, Guy},
	title       = {Generative {AI} as Seniority-Biased Technological Change: Evidence from {U.S.} R{\'e}sum{\'e} and Job Posting Data},
	institution = {SSRN},
	type        = {Working Paper},
	number      = {5425555},
	year        = {2025},
	month       = sep,
	doi         = {10.2139/ssrn.5425555}
}

@techreport{manning2025adaptable,
	author      = {Manning, Sam and Aguirre, Tom{\'a}s},
	title       = {How Adaptable Are {American} Workers to {AI}-Induced Job Displacement?},
	institution = {National Bureau of Economic Research},
	type        = {Working Paper},
	number      = {34705},
	year        = {2026},
	month       = jan,
	doi         = {10.3386/w34705}
}

@article{noy2023experimental,
	title={Experimental Evidence on the Productivity Effects of Generative {Artificial Intelligence}},
	author={Noy, Shakked and Zhang, Whitney},
	journal={Science},
	volume={381},
	number={6654},
	pages={187--192},
	year={2023},
	doi={10.1126/science.adh2586}
}

@misc{tomlinson2025working,
	title={Working with {AI}: Measuring the Applicability of Generative {AI} to Occupations},
	author={Tomlinson, Kiran and Jaffe, Sonia and Wang, Will and Counts, Scott and Suri, Siddharth},
	year={2025},
	howpublished={arXiv preprint arXiv:2507.07935},
	eprint={2507.07935},
	archiveprefix={arXiv}
}

@incollection{AcemogluAutor2011Handbook,
	title={Skills, Tasks and Technologies: Implications for Employment and Earnings},
	author={Acemoglu, Daron and Autor, David},
	booktitle={Handbook of Labor Economics},
	editor={Ashenfelter, Orley and Card, David},
	volume={4B},
	chapter={12},
	pages={1043--1171},
	year={2011},
	publisher={Elsevier},
	doi={10.1016/S0169-7218(11)02410-5}
}

@techreport{Webb2020SSRN,
	author      = {Webb, Michael},
	title       = {The Impact of {Artificial Intelligence} on the Labor Market},
	institution = {SSRN},
	type        = {Working Paper},
	number      = {3482150},
	year        = {2020},
	doi         = {10.2139/ssrn.3482150}
}

@techreport{NedelkoskaQuintini2018OECD,
	author      = {Nedelkoska, Ljubica and Quintini, Glenda},
	title       = {Automation, Skills Use and Training},
	institution = {OECD Publishing},
	type        = {OECD Social, Employment and Migration Working Papers},
	number      = {202},
	year        = {2018},
	doi         = {10.1787/2e2f4eea-en}
}

@techreport{Pizzinelli2023IMF,
	author      = {Pizzinelli, Carlo and Panton, Augustus J. and Tavares, Marina Mendes and Cazzaniga, Mauro and Li, Longji},
	title       = {Labor Market Exposure to {AI}: Cross-country Differences and Distributional Implications},
	institution = {International Monetary Fund},
	type        = {IMF Working Paper},
	number      = {WP/23/216},
	year        = {2023},
	doi         = {10.5089/9798400254802.001}
}

@article{silverMasteringGameGo2017,
	title={Mastering the Game of {Go} Without Human Knowledge},
	author={Silver, David and Schrittwieser, Julian and Simonyan, Karen and Antonoglou, Ioannis and Huang, Aja and Guez, Arthur and Hubert, Thomas and Baker, Lucas and Lai, Matthew and Bolton, Adrian and others},
	journal={Nature},
	volume={550},
	number={7676},
	pages={354--359},
	year={2017},
	doi={10.1038/nature24270}
}

@techreport{LashkariLiQiuThompson2026,
	author      = {Lashkari, Danial and Li, Wensu and Qiu, Christina and Thompson, Neil},
	title       = {{AI}, Scale, and Task-Based Theories of Automation},
	institution = {MIT Sloan School of Management},
	type        = {Working Paper},
	number      = {7381-26},
	year        = {2026},
	month       = jul,
	url         = {https://mitsloan.mit.edu/shared/ods/documents/?PublicationDocumentID=10949}
}

@misc{palazzolo2025coworkers,
	author       = {Palazzolo, Stephanie},
	title        = {How {Anthropic} and {OpenAI} Are Developing {AI} `Co-Workers'},
	howpublished = {The Information},
	month        = sep,
	year         = {2025}
}

@misc{roose2025mechanize,
	author       = {Roose, Kevin},
	title        = {This {A.I.} Company Wants to Take Your Job},
	howpublished = {The New York Times},
	month        = jun,
	year         = {2025}
}

@misc{paul2026meta,
	author       = {Paul, Katie and Horwitz, Jeff},
	title        = {Meta to Start Capturing Employee Mouse Movements, Keystrokes for {AI} Training Data},
	howpublished = {Reuters},
	month        = apr,
	year         = {2026}
}

\newpage
\appendix

\section{Extended Methodology}
\label{app:methods}

We use O*NET 30.0 (August 2025 release). The Task Ratings file contains multiple scale types per task (importance, relevance, frequency). We extract the four identifying columns (O*NET-SOC Code, Title, Task ID, Task) and deduplicate, yielding 17,951 unique occupation--task pairs across 894 occupations defined at the 8-digit O*NET-SOC level (e.g., 15-1252.00 for Software Developers). Separately, we extract the Importance scale ratings (1--5) from the same file for use in occupation-level aggregation (see below).

The scoring rubric (v4.2; full text in Appendix~\ref{app:prompt}) has three stages.

First, a \textbf{Physical Feasibility Gate} determines whether the task can be performed primarily through digital means. Tasks requiring substantial physical embodiment fail the gate and receive an RL index of 0 without further scoring. We treat physical embodiment as a binary gate rather than a continuous dimension because RL environments exist in software: if a task fundamentally requires a physical body, favorable scores on other dimensions do not make RL feasible.

Second, for tasks that pass the gate, a \textbf{Structured Task Reasoning} step requires the model to classify the task type (generative, analytical, interactive, procedural, or hybrid), identify the core output, name the verification bottleneck, describe tool requirements, and predict which dimension will be the binding constraint. This step anchors the model's reasoning before scoring begins. It also serves as an auditability check: when the predicted binding constraint does not match the actual lowest-scoring dimension, it flags potentially inconsistent reasoning.

Third, the model scores eight dimensions on a 1--10 integer scale (1 = RL infeasible, 10 = ideal for RL), grouped into five categories reflecting their role in the RL training pipeline:
\begin{itemize}[nosep]
    \item \textbf{Reward Signal} (D1: Verification Method Spectrum). Where the task falls between deterministic verification (code that compiles) and contested professional judgment (psychotherapy outcomes), incorporating advances in rubric-based RL and AI-as-Judge.
    \item \textbf{Prerequisite} (D2: Environment Simulability). Whether the task setting can be cheaply replicated digitally. Without a simulable environment, RL training is impractical.
    \item \textbf{MDP Structure} (D3: State Observability; D4: Task Variability; D5: Sequential Decision Depth). These determine how well the task maps onto a Markov Decision Process. Partial observability, high input variability, and deep sequential decision chains make learning harder but are surmountable.
    \item \textbf{Training Signal} (D6: Feedback Density and Decomposability). Sparse, delayed, or non-decomposable feedback slows learning and complicates credit assignment, but does not preclude RL.
    \item \textbf{Practical Barriers} (D7: Tool and Interface Accessibility; D8: Output Tangibility and Gradeability). Whether the task can be performed through programmatic interfaces rather than GUIs (D7), and whether it produces a concrete, inspectable artifact that can be graded independently of the process that created it (D8).
\end{itemize}

All eight dimensions receive equal weight (1/8 each). We adopt equal weighting because each dimension captures a conceptually distinct aspect of RL feasibility, and no single dimension dominates once the physical gate has been passed. A task that is perfectly verifiable but non-simulable is no more RL-amenable than one that is perfectly simulable but non-verifiable. Equal weighting also avoids arbitrary expert prioritization among dimensions whose relative importance may vary across task types.

Each of the 17,951 occupation--task pairs is scored in a single API request to the OpenRouter API (\texttt{https://openrouter.ai/api/v1}). The request contains the full rubric with the occupation title and task description substituted into the prompt placeholders. We use Gemini 3.8 Flash (\texttt{google/gemini-3.8-flash}) with the following settings: temperature 0 (for reproducibility), maximum output tokens 4,000, reasoning effort \texttt{medium}, and structured JSON output enforced via \texttt{response\_format: \{type: json\_object\}}.

The model first evaluates the Physical Feasibility Gate, returning a binary pass/fail with a 2--3 sentence justification. Tasks that fail receive an RL index of 0 and no dimension scores. For tasks that pass, the model performs the structured task reasoning step, then scores each of the eight dimensions. For each dimension, the model writes a 2--3 sentence justification \textit{before} assigning its integer score. This reason-then-score design forces chain-of-thought reasoning and reduces default or arbitrary ratings. The prompt also includes an explicit instruction to resist central tendency bias. We compute the RL Feasibility Index externally from the returned scores (Equation~\ref{eq:rl_index}) rather than asking the model to compute it, eliminating arithmetic errors.

We process all tasks concurrently using 50 parallel requests via asynchronous HTTP (\texttt{asyncio} + \texttt{aiohttp}). Failed requests are attempted up to 3 times with exponential backoff ($2^{\text{attempt}}$ seconds); rate-limited responses (HTTP 429) respect the server's \texttt{Retry-After} header. Each request times out after 120 seconds. The Gemini 3.8 Flash output contains 17,893 valid task scores. Eight tasks have no recorded gate decision, and 50 tasks pass the gate but lack a complete set of dimension scores. These 58 tasks are excluded from the analysis rather than assigned a zero. GPT-5.6 Luna and GLM-5.3 have 4 and 24 missing task scores, respectively.

Scores are conditioned on the occupation as well as the task text. The same task statement can describe very different work depending on the occupation. ``Prepare reports'' for a Statistical Assistant involves formatting numeric tables (high RL feasibility), while the same task for a Chief Sustainability Officer involves synthesizing qualitative evidence and stakeholder input (low RL feasibility). The prompt instructs the model to account for the complexity, stakes, expertise, and autonomy implied by the occupation context.

We aggregate task-level scores to occupations using importance-weighted averaging. The weight for task $i$ in occupation $j$ is $w_{ij} = \text{Imp}_{ij} / \sum_{k \in T_j} \text{Imp}_{kj}$, where $\text{Imp}_{ij}$ is the O*NET importance rating (1--5 scale) and $T_j$ is the set of tasks with valid scores in occupation $j$. We renormalize the weights over these tasks. Core duties therefore contribute more than peripheral tasks. In practice, the importance-weighted and unweighted occupation-level means correlate at 0.9995, indicating that the weighting has minimal effect. We report the weighted version throughout.

We merge our occupation-level index with the public task-level labels from \citet{eloundou2024gpts}.\footnote{Eloundou et al.\ use O*NET~27.2; we use O*NET~30.0, so the set of occupations and tasks differs slightly. We re-aggregate their task-level $\beta$ scores to the occupation level using O*NET~30.0 importance weights, rather than their original weighting scheme (weight of 1 for core tasks, 0.5 for supplementary tasks). This ensures both indices are aggregated on the same basis.} Their $\beta$ score is a GPT-4-based exposure measure taking three values: 0 (not exposed), 0.5 (exposed with additional software), and 1 (exposed to standalone LLMs, meaning an LLM could reduce task completion time by at least 50\%). We focus on $\beta$ because it is the most widely cited of their measures and the most directly comparable to our index.

We join the Eloundou task-level data with O*NET~30.0 importance ratings by occupation and task identifier, compute importance-weighted means of $\beta$ within each occupation, and merge with our occupation-level RL index by O*NET-SOC code. We report Pearson and Spearman correlations at the occupation level. The quadrant comparison classifies occupations relative to the sample median of each index. Occupations above the $\beta$ median but below the RL median have high general AI exposure and low RL feasibility; those in the opposite quadrant have high RL feasibility and low general AI exposure.

\section{Full Scoring Prompt}
\label{app:prompt}

The following is the complete prompt (v4.2) used to score each occupation--task pair. The placeholders \texttt{\{\{OCCUPATION\}\}} and \texttt{\{\{TASK\}\}} are replaced with the specific occupation title and task description for each item.

\bigskip
\hrule
\bigskip

\small
\begin{quote}

    \textbf{Context}

    You are an expert in reinforcement learning (RL), labour economics, and occupational task analysis. You are helping construct a \textbf{Reinforcement Learning Feasibility Index} --- a measure of how feasible it is to create an RL environment for a given occupational task, and therefore how exposed that task is to automation through advances in RL-based training (including RL post-training of large language models).

    This index is analogous to Eloundou et al.'s (2024, \textit{Science}) AI Exposure Index, but focused specifically on reinforcement learning. For each task, we ask: \textit{How easy would it be to define a well-specified RL environment --- with a reward signal, simulable environment, and verifiable outcomes --- in which an agent could learn to perform this task?}

    The more feasible it is to build such an environment, the more ``RL-exposed'' the task is, because:
    \begin{itemize}[nosep]
        \item RL post-training (RLHF, RLAIF, RLVR) is a primary driver of LLM capability gains
        \item Tasks with verifiable outcomes and simulable environments are precisely those where RL training is most effective
        \item Advances in RL will disproportionately improve AI performance on tasks that score highly on this index
    \end{itemize}

    This version incorporates advances in \textbf{Rubric-Based RL} (e.g., Rubric-ARM, RLVR, OpenRubrics) and \textbf{Process Reward Models}, which demonstrate that subjective tasks are feasible for RL if they can be decomposed into verifiable criteria.

    \textbf{Crucial Assumption (The Agent's Starting Point):} Assume the RL agent is \textit{not} learning from scratch (\textit{tabula rasa}). Instead, assume we are fine-tuning a highly capable pre-trained foundation model (e.g., via RLHF, RLAIF, or RLVR). The agent already possesses a broad baseline understanding of language, code, and general world knowledge. Your evaluation should focus on the feasibility of the RL loop required to align, specialize, and verify the model for this specific occupational task.

    \textbf{Temporal Frame}

    Score based on feasibility using methods that are currently available or plausibly achievable within a 5-year horizon. Include the trajectory of AI-as-Judge, Process Reward Models, and Synthetic Data Generation. Do not assume speculative breakthroughs with no current research basis. When a score depends on projected rather than current capabilities, note this in the justification.

    \textbf{Input}

    You will be given an \textbf{occupation title} and a \textbf{task description}. You must score the RL feasibility of performing that specific task \textbf{in the context of that specific occupation}.

    \textbf{This is critical}: the same task text can have very different RL feasibility depending on the occupation.

    \textbf{Step 1: Physical Feasibility Gate}

    Before scoring any dimensions, first determine whether the task \textbf{requires substantial physical interaction with the material world}.

    \textit{Decision rule}: \textbf{Pass}: The task can be performed primarily through digital means. Proceed to Step 2. \textbf{Fail}: The task irreducibly requires physical embodiment. Do not proceed further.

    \textbf{Step 2: Structured Task Reasoning}

    Before scoring any dimensions, reason about the task holistically: (1) Task type (generative, analytical, interactive, procedural, or hybrid). (2) Core output. (3) Verification bottleneck. (4) Tool requirements. (5) Binding constraint (predicted lowest-scoring dimension).

    \textbf{Step 3: Score the 8 Dimensions}

    Score it on each of the \textbf{8 dimensions} below. Each dimension must be scored as a \textbf{strict integer on a 10-point scale (1--10)}.

    \textbf{Avoid central tendency bias}: Confidently use the extreme ends of the scale (1--3 and 8--10) when the task characteristics warrant it.

    For each dimension, provide:
    \begin{enumerate}[nosep]
        \item A \textbf{2--3 sentence justification} explaining the score. \textbf{Write the justification first, then assign the score.}
        \item An integer score (1--10)
    \end{enumerate}

    \textbf{D1: Verification Method Spectrum.} Where does this task fall on the spectrum from deterministic verification to contested professional judgment? Scores: 1 (requires rare experts who disagree; no inspectable artifact) to 10 (fully deterministic programmatic check).

    \textbf{D2: Environment Simulability.} How faithfully and cheaply can the task environment be simulated digitally? Scores: 1 (requires live markets, real humans with genuine stakes) to 10 (natively digital, trivially cheap to replicate).

    \textbf{D3: State Observability \& Context.} To what extent is all relevant information available in a structured, digital format? Scores: 1 (critical information is tacit or embodied) to 10 (perfect digital observability).

    \textbf{D4: Task Variability \& Knowledge Breadth.} How varied are the inputs across task instances, and how broad is the required expertise? Scores: 1 (extreme variability, every instance unique) to 10 (zero variability, structurally identical instances).

    \textbf{D5: Sequential Decision Depth.} How many counterfactual-sensitive decisions must be made in sequence? Scores: 1 (extreme depth, dozens of contingent decisions) to 10 (single-step, one input, one output).

    \textbf{D6: Feedback Density \& Decomposability.} How frequently and specifically does the agent receive performance signals, and how decomposable are they? Scores: 1 (rare, delayed, holistic, non-decomposable) to 10 (continuous, immediate, per-step diagnostic).

    \textbf{D7: Tool \& Interface Accessibility.} Can the task be performed through CLI/API/MCP, or does it require GUI interaction? Scores: 1 (complex proprietary GUIs, no scripting) to 10 (natively machine-interfaced or pure text/code generation).

    \textbf{D8: Output Tangibility \& Gradeability.} Does the task produce a concrete, inspectable artifact that can be evaluated independently of the process? Scores: 1 (no tangible output, quality in ongoing process/relationship) to 10 (perfectly tangible, self-contained artifact).

    \textbf{Output Format}

    Return a JSON object with: occupation, task, physical\_feasibility (justification and pass boolean), task\_reasoning (task type, core output, verification bottleneck, tool requirements, binding constraint), and dimensions (each with justification and integer score 1--10, or null if the task failed the gate). Do not compute or include the RL Feasibility Index.

    \bigskip
    \textbf{Occupation}: \texttt{\{\{OCCUPATION\}\}}

    \textbf{Task}: \texttt{\{\{TASK\}\}}

\end{quote}

\bigskip
\hrule

\normalsize

\section{Additional Descriptives for RL Index}
\label{app:descriptives}

Table~\ref{tab:desc_summary} reports summary statistics at the task and occupation levels. Among the 17,893 scored tasks, 43.9\% fail the physical feasibility gate and receive a score of zero; the 10,045 scored gate-passing tasks have a conditional mean of 52.0. Aggregation to occupations compresses the distribution (SD is 18.5 at the occupation level compared with 28.9 at the task level) and reduces the concentration at zero, since most occupations contain a mix of physical and non-physical tasks. Forty-seven occupations retain a score of zero.

\begin{table}[!htbp]
\centering
\caption{Summary statistics for the RL feasibility index. Model: Gemini 3.8 Flash}
\label{tab:desc_summary}
\small
\begin{tabular}{lrrrrrrrr}
\toprule
\textbf{Level} & \textbf{N} & \textbf{Mean} & \textbf{SD} & \textbf{P10} & \textbf{P25} & \textbf{Median} & \textbf{P75} & \textbf{P90} \\
\midrule
Task & 17,893 & 29.2 & 28.9 & 0.0 & 0.0 & 27.8 & 56.9 & 69.4 \\
Occupation (wtd.) &    894 & 29.0 & 18.5 & 4.6 & 13.1 & 28.9 & 42.0 & 54.2 \\
\bottomrule
\end{tabular}
\caption*{\footnotesize \textit{Notes:} 43.9\% of scored tasks fail the physical feasibility gate and score 0. Occupation-level scores are importance-weighted means of task scores. Tasks without a score: 58. Importance weights are renormalized over valid scores within each occupation.}
\end{table}

\begin{figure}[h!]
    \centering
    \caption{Task-level RL Feasibility Index distribution.}
    \label{fig:desc_task_dist}
    \includegraphics[width=0.85\textwidth]{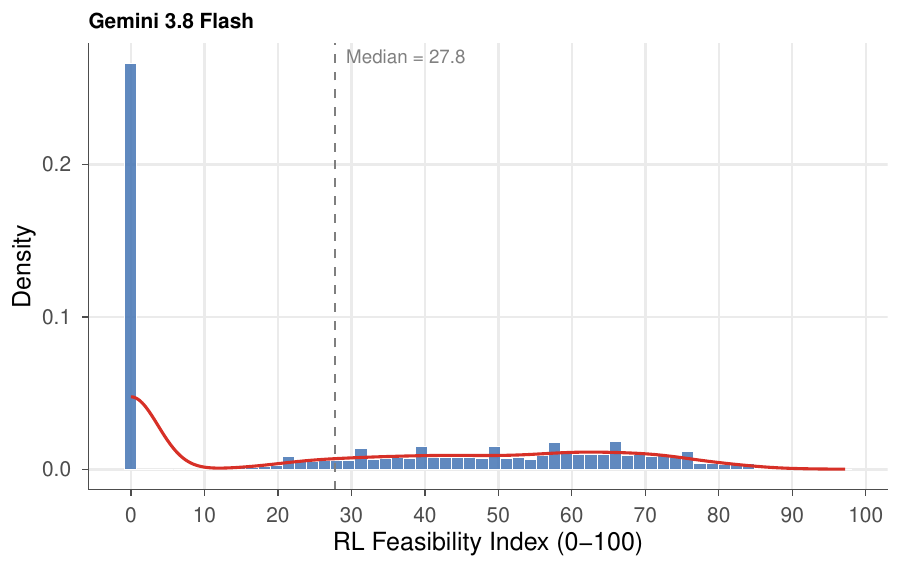}
\caption*{\footnotesize \textit{Notes:} Distribution of RL Feasibility Index scores across 17,893 scored O*NET 30.0 tasks. Scores are rescaled to 0--100 (Equation~\ref{eq:rl_index}). The spike at zero reflects the 43.9\% of scored tasks that fail the physical feasibility gate. Among scored gate-passing tasks ($N = 10{,}045$), the conditional mean is 52.0.}
\end{figure}

\begin{figure}[h!]
    \centering
    \caption{Occupation-level RL Feasibility Index distribution (importance-weighted means).}
    \label{fig:desc_occ_dist}
    \includegraphics[width=0.85\textwidth]{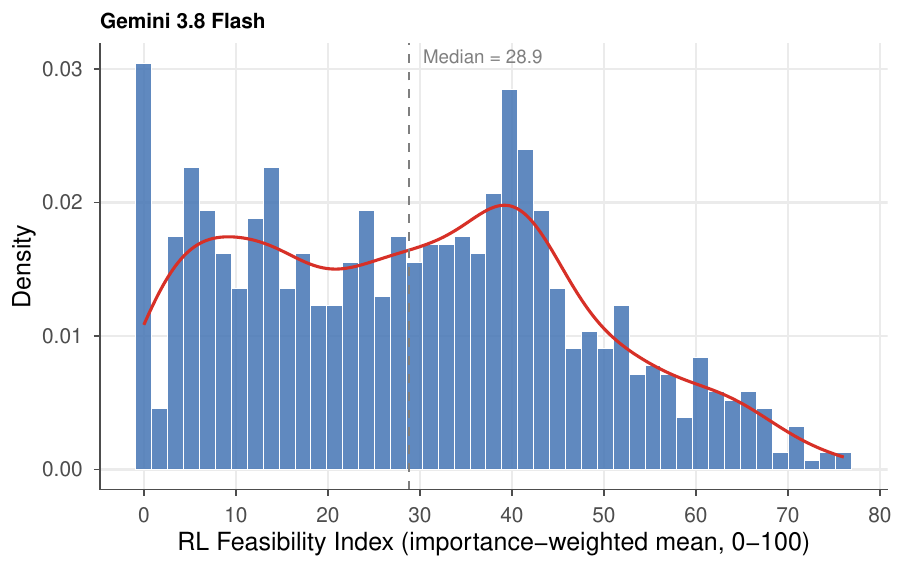}
    \caption*{\footnotesize \textit{Notes:} Distribution of occupation-level RL Feasibility Index scores ($N = 894$). Each score is the importance-weighted mean of constituent task scores (Equation~\ref{eq:rl_index}), using O*NET 30.0 importance ratings. The occupation-level mean is 29.0 (SD 18.5). Most occupations contain a mix of gate-passing and gate-failing tasks; 47 occupations score zero.}
\end{figure}

Figure~\ref{fig:desc_task_dist} shows the task-level distribution. The spike at zero reflects gate failures; scored gate-passing tasks have a mean of 52.0. Figure~\ref{fig:desc_occ_dist} shows the more compressed occupation-level distribution, with a mean of 29.0.

Table~\ref{tab:desc_dimensions} reports means and dispersion for the eight dimensions among tasks with complete scores that pass the gate. Output Tangibility (D8, mean 6.6) and Tool Accessibility (D7, mean 6.6) score highest. Task Variability (D4, mean 4.4) and Verification Method (D1, mean 5.2) score lowest, suggesting that input diversity and difficulties in verifying outcomes are common barriers to RL training.

\begin{table}[!htbp]
\centering
\caption{Dimension score summary statistics (gate-passing tasks). Model: Gemini 3.8 Flash}
\label{tab:desc_dimensions}
\small
\begin{tabular}{lrrrrr}
\toprule
\textbf{Dimension} & \textbf{Mean} & \textbf{SD} & \textbf{P25} & \textbf{Median} & \textbf{P75} \\
\midrule
D1: Verification Method & 5.2 & 2.1 & 3 & 5 & 7 \\
D2: Environment Simulability & 5.6 & 2.1 & 4 & 5 & 8 \\
D3: State Observability & 5.8 & 1.8 & 4 & 6 & 7 \\
D4: Task Variability & 4.4 & 1.7 & 3 & 4 & 6 \\
D5: Sequential Decision Depth & 5.4 & 1.7 & 4 & 5 & 7 \\
D6: Feedback Density & 5.9 & 2.0 & 4 & 6 & 7 \\
D7: Tool Accessibility & 6.6 & 1.8 & 5 & 7 & 8 \\
D8: Output Tangibility & 6.6 & 2.5 & 5 & 8 & 9 \\
\bottomrule
\end{tabular}
\caption*{\footnotesize \textit{Notes:} Each dimension is scored on a 1--10 integer scale. Statistics over the 10,045 tasks that pass the physical feasibility gate.}
\end{table}

\begin{figure}[h!]
    \centering
    \caption{Pairwise correlations among the eight RL feasibility dimensions.}
    \label{fig:desc_dim_corr}
    \includegraphics[width=0.75\textwidth]{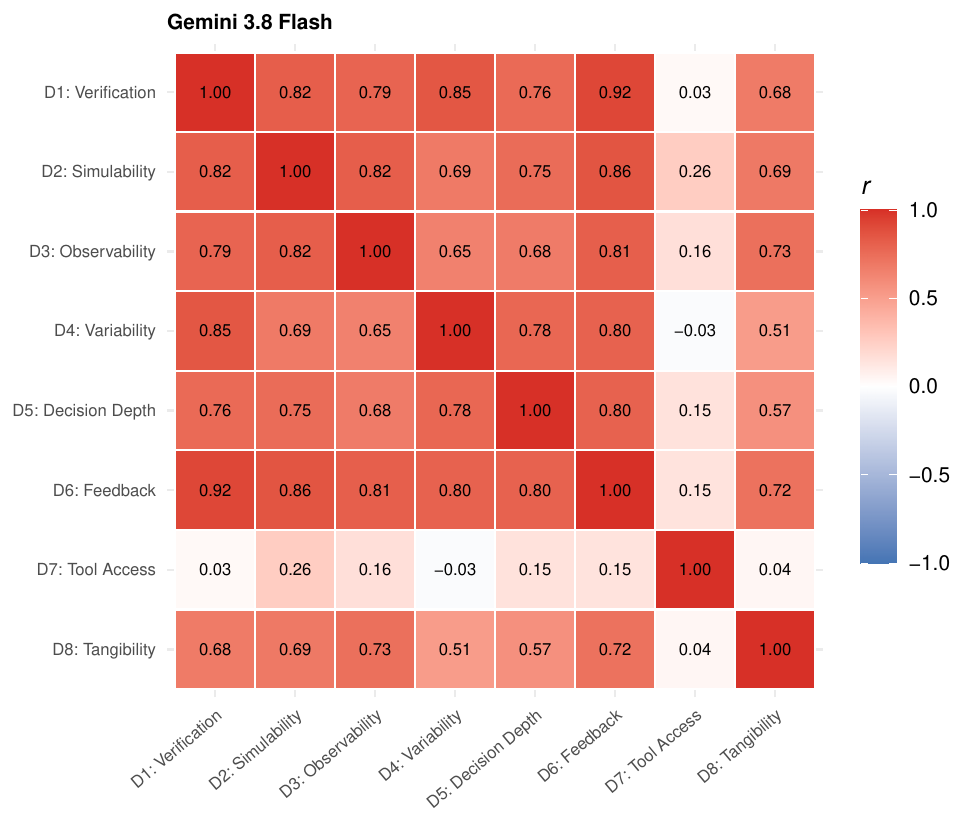}
    \caption*{\footnotesize \textit{Notes:} Pairwise Pearson correlations among the eight RL feasibility dimension scores, computed over 10,045 scored tasks that pass the physical feasibility gate. Each dimension is scored on a 1--10 integer scale (1 = RL infeasible, 10 = ideal for RL).}
\end{figure}

\subsection{Principal Component Analysis}
\label{app:pca}

We conduct a principal component analysis (PCA) on the eight dimension scores (standardized, correlation matrix) for gate-passing tasks with complete dimension scores to assess the dimensionality of the RL feasibility construct. Table~\ref{tab:pca_variance} reports eigenvalues and variance explained. The first principal component has an eigenvalue of 5.521 and accounts for 69.0\% of total variance. The Kaiser criterion retains PC1 and PC2, and Horn's parallel analysis (1,000 simulations) also retains two components. The dominance of PC1 is consistent with the high Cronbach's $\alpha$ of 0.919: the eight dimensions, while conceptually distinct, share a strong common factor, alongside a second component dominated by tool accessibility.

\begin{table}[!htbp]
\centering
\caption{PCA variance explained. Model: Gemini 3.8 Flash}
\label{tab:pca_variance}
\small
\begin{tabular}{lrrr}
\toprule
\textbf{Component} & \textbf{Eigenvalue} & \textbf{Var. Explained (\%)} & \textbf{Cumulative (\%)} \\
\midrule
PC1 & 5.521 & 69.0 & 69.0 \\
PC2 & 1.054 & 13.2 & 82.2 \\
PC3 & 0.583 & 7.3 & 89.5 \\
PC4 & 0.266 & 3.3 & 92.8 \\
PC5 & 0.218 & 2.7 & 95.5 \\
PC6 & 0.174 & 2.2 & 97.7 \\
PC7 & 0.120 & 1.5 & 99.2 \\
PC8 & 0.065 & 0.8 & 100.0 \\
\bottomrule
\end{tabular}
\caption*{\footnotesize \textit{Notes:} PCA on the correlation matrix of the eight standardised dimension scores ($N = 10,045$ gate-passing tasks with complete scores). Cronbach's alpha (standardised) = 0.919. Horn's parallel analysis (1,000 simulations) retains 2 components.}
\end{table}

\begin{figure}[h!]
    \centering
    \caption{Scree plot with parallel analysis.}
    \label{fig:pca_scree}
    \includegraphics[width=0.75\textwidth]{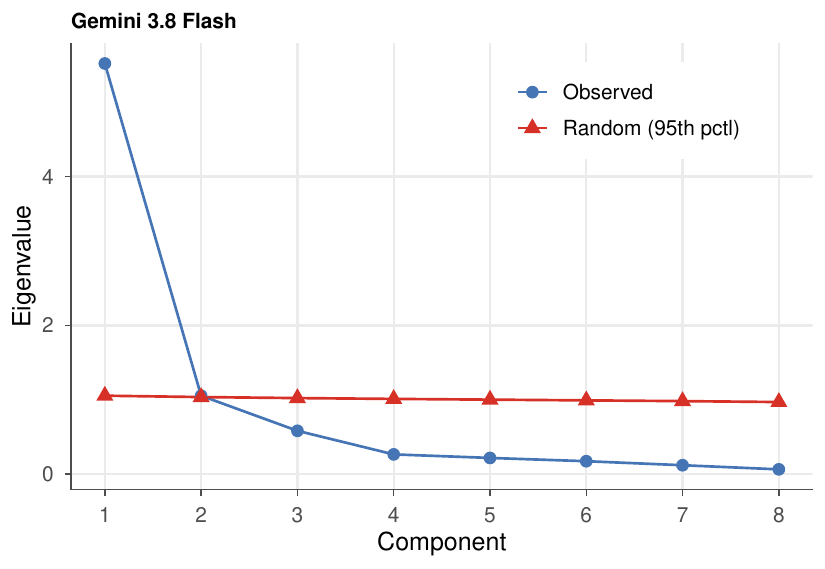}
    \caption*{\footnotesize \textit{Notes:} Solid line shows eigenvalues from PCA on the correlation matrix of eight standardized dimension scores ($N = 10{,}045$ scored gate-passing tasks). Dashed line shows 95th-percentile eigenvalues from Horn's parallel analysis (1,000 simulations of random data with the same dimensions). PC1 and PC2 exceed the random threshold.}
\end{figure}

Figure~\ref{fig:pca_scree} shows the scree plot alongside the 95th-percentile eigenvalues from parallel analysis. PC1 and PC2 exceed the random threshold, indicating a dominant general feasibility factor alongside a distinct second component.

\begin{figure}[h!]
    \centering
    \caption{PCA component weights heatmap.}
    \label{fig:pca_loadings}
    \includegraphics[width=0.75\textwidth]{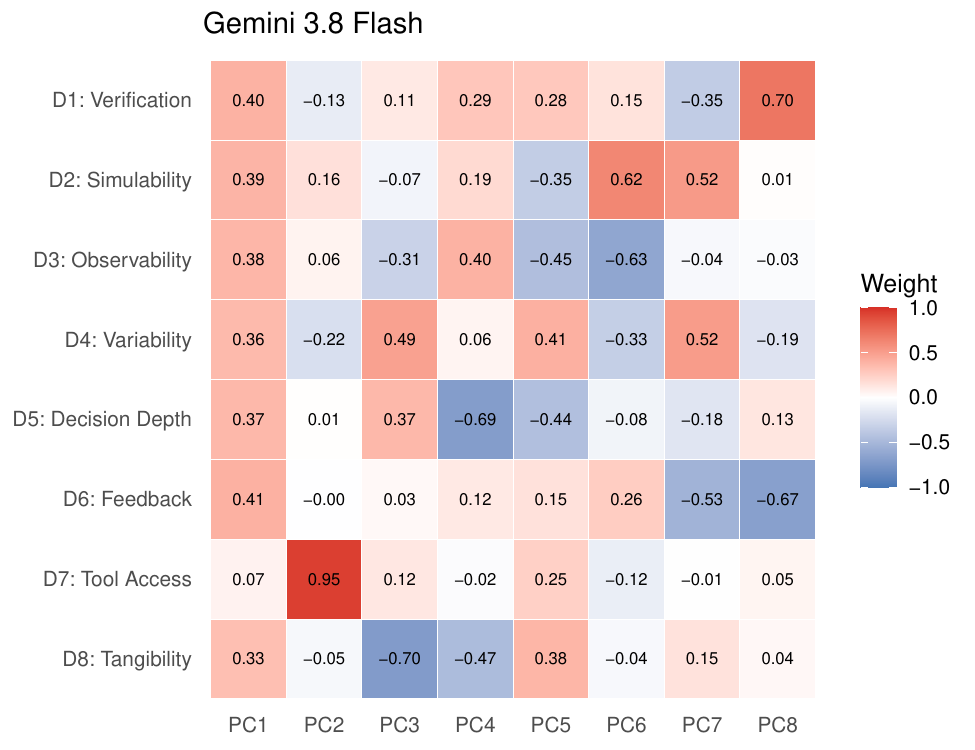}
    \caption*{\footnotesize \textit{Notes:} Component weights from PCA on the correlation matrix of eight standardized dimension scores ($N = 10{,}045$ scored gate-passing tasks). PC1 explains 69.0\% of variance; PC2 explains 13.2\%. Weights are the eigenvector coefficients used to construct each component from the standardized dimensions.}
\end{figure}

Figure~\ref{fig:pca_loadings} displays the component weights. PC1 loads positively on all eight dimensions, with the highest weights on feedback (0.41), verification (0.40), and simulability (0.39). PC2 loads almost entirely on tool access (0.95), isolating a dimension that is comparatively distinct from the other seven.

\begin{figure}[h!]
    \centering
    \caption{PCA biplot (PC1 vs.\ PC2) with dimension weight vectors.}
    \label{fig:pca_biplot}
    \includegraphics[width=0.75\textwidth]{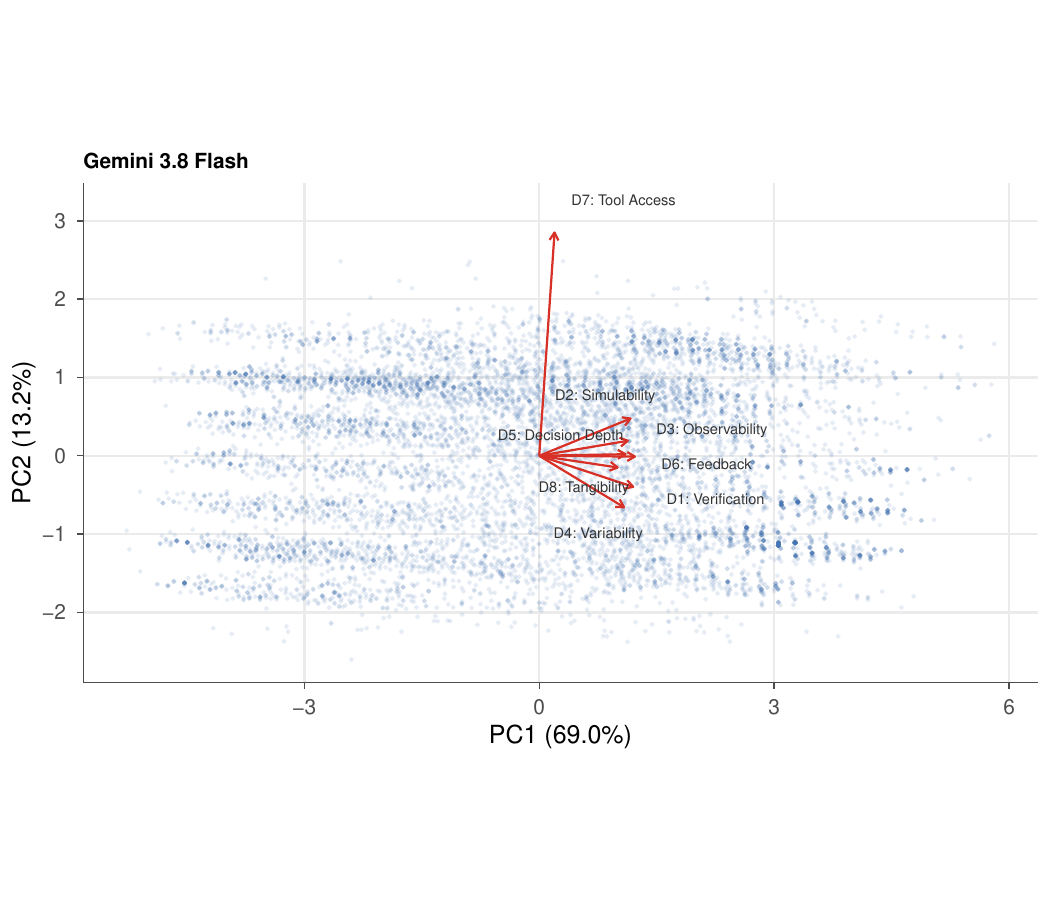}
    \caption*{\footnotesize \textit{Notes:} Each blue point is one scored gate-passing task ($N = 10{,}045$) projected onto PC1 and PC2. Arrows show dimension weight vectors scaled for visibility. Arrow direction indicates which dimensions drive variation along each component. PC1 (horizontal) explains 69.0\% of variance; PC2 (vertical) explains 13.2\%.}
\end{figure}

Figure~\ref{fig:pca_biplot} presents a biplot of task scores on PC1 and PC2, with dimension weight vectors overlaid. Most vectors cluster tightly along the PC1 axis, while tool access points almost exclusively along PC2. The cloud of task scores is elongated along PC1, visually confirming that a single factor accounts for most of the variation in RL feasibility.

\clearpage
\section{Comparison of Annotation Models}
\label{app:model_comparison}

We compare the exposure measures produced by Gemini 3.8 Flash, GPT-5.6 Luna, and GLM-5.3 using the same rubric and the same 17,951 O*NET occupation--task pairs. Table~\ref{tab:model_comparison} reports coverage and summary statistics. There are 58, 4, and 24 missing task scores, respectively. For each model, we exclude missing scores and renormalize task-importance weights within occupations; all 894 occupations retain an exposure score. Missing scores are distinct from gate failures, which receive a zero.

The occupation-level means are 29.0 for Gemini 3.8 Flash, 29.3 for GPT-5.6 Luna, and 33.1 for GLM-5.3. The higher GLM-5.3 mean reflects more generous scores among tasks that pass the gate: their conditional mean is 59.6, compared with 52.0 and 52.1 for the other models. The share of scored tasks assigned zero is similar across models, at 42.8--43.9\%. Model choice therefore matters for exposure levels even when the physical gate classifies a similar share of tasks as infeasible.

\begin{table}[!htbp]
\centering
\footnotesize
\caption{Coverage and distribution of the three exposure measures}
\label{tab:model_comparison}
\begin{tabular}{lrrr}
\toprule
Statistic & Gemini 3.8 Flash & GPT-5.6 Luna & GLM-5.3 \\
\midrule
Tasks with valid scores & 17,893 & 17,947 & 17,927 \\
Missing task scores & 58 & 4 & 24 \\
Scored tasks at zero (\%) & 43.9 & 42.8 & 43.5 \\
Task mean & 29.2 & 29.8 & 33.6 \\
Task SD & 28.9 & 28.0 & 31.9 \\
Gate-passing task mean & 52.0 & 52.1 & 59.6 \\
Occupation mean & 29.0 & 29.3 & 33.1 \\
Occupation SD & 18.5 & 18.2 & 21.5 \\
Occupations at zero & 47 & 46 & 50 \\
\bottomrule
\end{tabular}
\caption*{\footnotesize \textit{Notes:} Task statistics use all valid scores for each model. Zero shares use scored tasks as the denominator. Occupation means use O*NET task-importance weights renormalized over valid scores. Each model covers 894 occupations. Index scale: 0--100.}
\end{table}

\begin{figure}[!htbp]
\centering
\caption{Occupation-level exposure across annotation models.}
\label{fig:model_pairs}
\includegraphics[width=\textwidth]{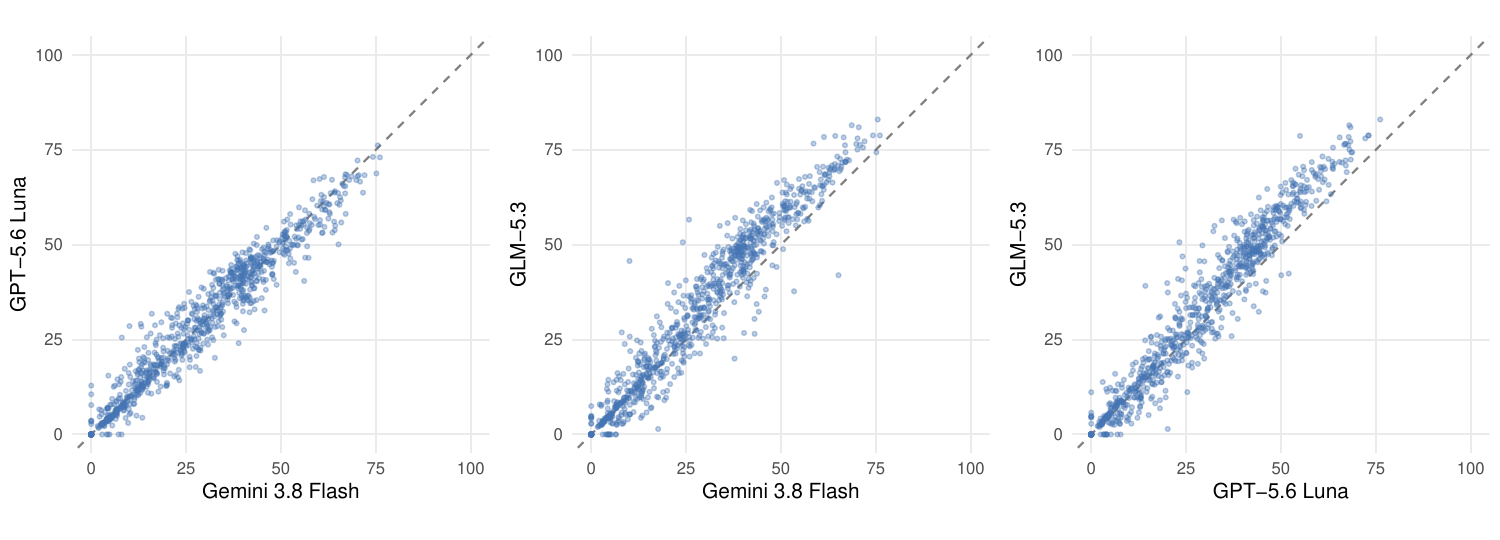}
\caption*{\footnotesize \textit{Notes:} Each point is one of 894 occupations. Scores are importance-weighted means over each model's valid task scores. Dashed lines mark equality. All axes use the same 0--100 scale.}
\end{figure}

Table~\ref{tab:model_pairs} compares all three pairs of models. Task-level correlations and gate agreement use the common sample of 17,866 tasks with valid scores from all three annotators. Occupation-level comparisons use all 894 occupations, with each model's available task scores. Pearson correlations range from 0.965 to 0.971 at the occupation level and from 0.868 to 0.898 at the task level; occupation rank correlations range from 0.964 to 0.970. Figure~\ref{fig:model_pairs} plots the occupation-level comparisons. The strong agreement in rankings is consistent with the common ordering of the most and least exposed major occupation groups, although individual occupations can receive different scores.

\begin{table}[!htbp]
\centering
\footnotesize
\caption{Pairwise agreement across annotation models}
\label{tab:model_pairs}
\begin{tabular}{lrrr}
\toprule
Model pair & Task Pearson & Occ. Pearson & Occ. Spearman \\
\midrule
Gemini 3.8 Flash / GPT-5.6 Luna & 0.898 & 0.971 & 0.970 \\
Gate agreement (\%) & 95.3 &  &  \\
Gemini 3.8 Flash / GLM-5.3 & 0.870 & 0.965 & 0.964 \\
Gate agreement (\%) & 94.1 &  &  \\
GPT-5.6 Luna / GLM-5.3 & 0.868 & 0.970 & 0.969 \\
Gate agreement (\%) & 93.6 &  &  \\
\bottomrule
\end{tabular}
\caption*{\footnotesize \textit{Notes:} Task correlations and gate agreement use 17,866 tasks with valid scores from all three models. Occupation correlations use all 894 occupations and each model's available scores. Pearson correlations measure score agreement; Spearman correlations measure rank agreement.}
\end{table}

Table~\ref{tab:model_results} compares the relationship with existing AI exposure measures. The correlation with Eloundou et al.'s $\beta$ measure is high under all three models (0.912--0.921), but falls to 0.379--0.434 when both measures are restricted to tasks passing each model's physical gate. These restricted samples differ across models, so the comparison reflects both scoring differences and changes in the task set. Correlations with the human-annotated $\beta$ measure are also high (0.854--0.890). The broad relationship with general AI exposure is therefore robust to annotator choice.

\begin{table}[!htbp]
\centering
\footnotesize
\caption{Comparison of external correlations}
\label{tab:model_results}
\begin{tabular}{lrrr}
\toprule
Statistic & Gemini 3.8 Flash & GPT-5.6 Luna & GLM-5.3 \\
\midrule
Pearson correlation with $\beta$ & 0.912 & 0.916 & 0.921 \\
Spearman correlation with $\beta$ & 0.922 & 0.922 & 0.932 \\
Pearson correlation with human $\beta$ & 0.854 & 0.866 & 0.890 \\
Gate-passing matched tasks & 10,011 & 10,238 & 10,087 \\
Gate-passing occupations & 847 & 848 & 844 \\
Pearson correlation with $\beta$, gated & 0.379 & 0.415 & 0.434 \\
Spearman correlation with $\beta$, gated & 0.413 & 0.438 & 0.457 \\
\bottomrule
\end{tabular}
\caption*{\footnotesize \textit{Notes:} Full-sample correlations use 894 occupations. Gated correlations restrict both indices to tasks with valid RL scores that pass each model's gate before aggregation.}
\end{table}

\clearpage
\clearpage
\section{Comparison with Alternative AI Exposure Measures}
\label{app:eloundou_alternatives}

\begin{figure}[!htbp]
    \centering
    \caption{Correlations with alternative AI exposure measures, by sample.}
    \label{fig:eloundou_alternatives}
    \includegraphics[width=0.9\textwidth]{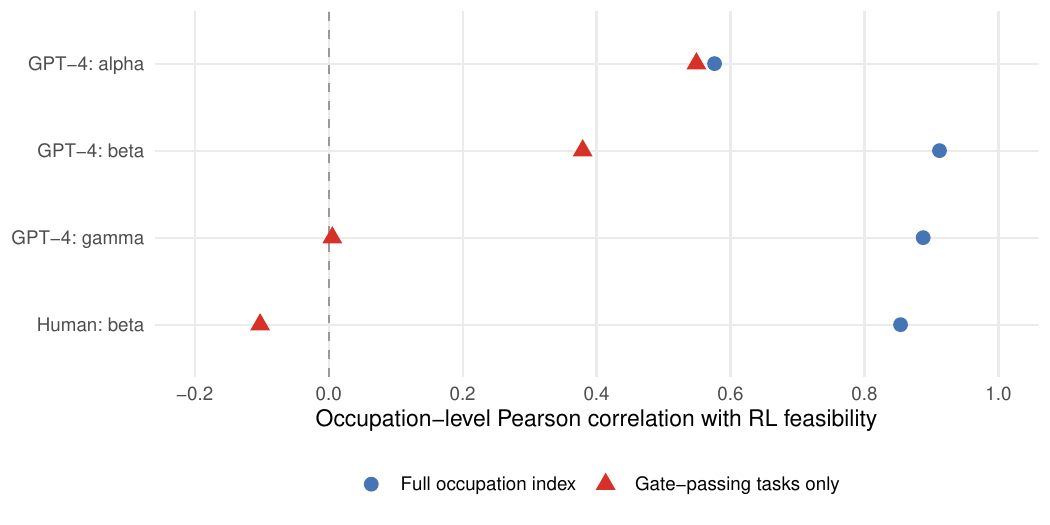}
    \caption*{\footnotesize \textit{Notes:} Points show occupation-level Pearson correlations between the Gemini 3.8 Flash RL Feasibility Index and each exposure measure. The full comparison covers 894 occupations; the gate-passing comparison covers 847 occupations and re-aggregates both indices over the same 10,011 matched tasks. The dashed line marks zero correlation. Points are descriptive correlations, with no confidence intervals plotted.}
\end{figure}

This appendix reports the comparison between our Gemini 3.8 Flash index and the four exposure measures from \citet{eloundou2024gpts} that Section~\ref{sec:comparison} defines. Table~\ref{tab:eloundou_alternatives} reports Pearson and Spearman correlations for two samples, and Figure~\ref{fig:eloundou_alternatives} plots the Pearson correlations. The full comparison covers 894 occupations and uses each index's available task scores. The restricted comparison uses the 10,011 matched tasks with valid Gemini 3.8 Flash scores that pass the physical feasibility gate. Both indices are re-aggregated over these tasks, which leaves 847 occupations. Missing task scores and unmatched importance weights are excluded from the relevant averages, with weights renormalized over the available values. The restriction therefore changes both the task composition within occupations and the set of occupations represented.

\begin{table}[!htbp]
\centering
\small
\caption{Correlations between Gemini 3.8 Flash RL feasibility and alternative AI exposure measures}
\label{tab:eloundou_alternatives}
\begin{tabular}{lrrrr}
\toprule
& \multicolumn{2}{c}{Full occupation index} & \multicolumn{2}{c}{Gate-passing tasks only} \\
\cmidrule(lr){2-3}\cmidrule(lr){4-5}
Exposure measure & Pearson & Spearman & Pearson & Spearman \\
\midrule
$\alpha$ (GPT-4) & 0.576 & 0.543 & 0.549 & 0.485 \\
$\beta$ (GPT-4) & 0.912 & 0.922 & 0.379 & 0.413 \\
$\gamma$ (GPT-4) & 0.888 & 0.905 & 0.005 & 0.154 \\
$\beta$ (human) & 0.854 & 0.872 & $-0.102$ & $-0.060$ \\
\midrule
Occupations & \multicolumn{2}{c}{894} & \multicolumn{2}{c}{847} \\
\bottomrule
\end{tabular}
\caption*{\footnotesize \textit{Notes:} All correlations are at the occupation level. The RL index uses Gemini 3.8 Flash scores. Full indices are separately aggregated using O*NET 30.0 task-importance weights. In the restricted sample, both indices are re-aggregated over the same 10,011 matched tasks with valid RL scores that pass the physical gate. The exposure measures are $\alpha=E1$, $\beta=E1+0.5E2$, and $\gamma=E1+E2$, where $E1$ denotes exposure to standalone LLMs and $E2$ exposure requiring additional software. The human measure applies the $\beta$ weights to human annotations. Missing scores or weights are excluded from the relevant averages, with weights renormalized over available values.}
\end{table}

\begin{figure}[!htbp]
    \centering
    \caption{Occupation-level RL feasibility vs.\ general AI exposure ($\alpha$).}
    \label{fig:quadrant_alpha}
    \includegraphics[width=0.8\textwidth]{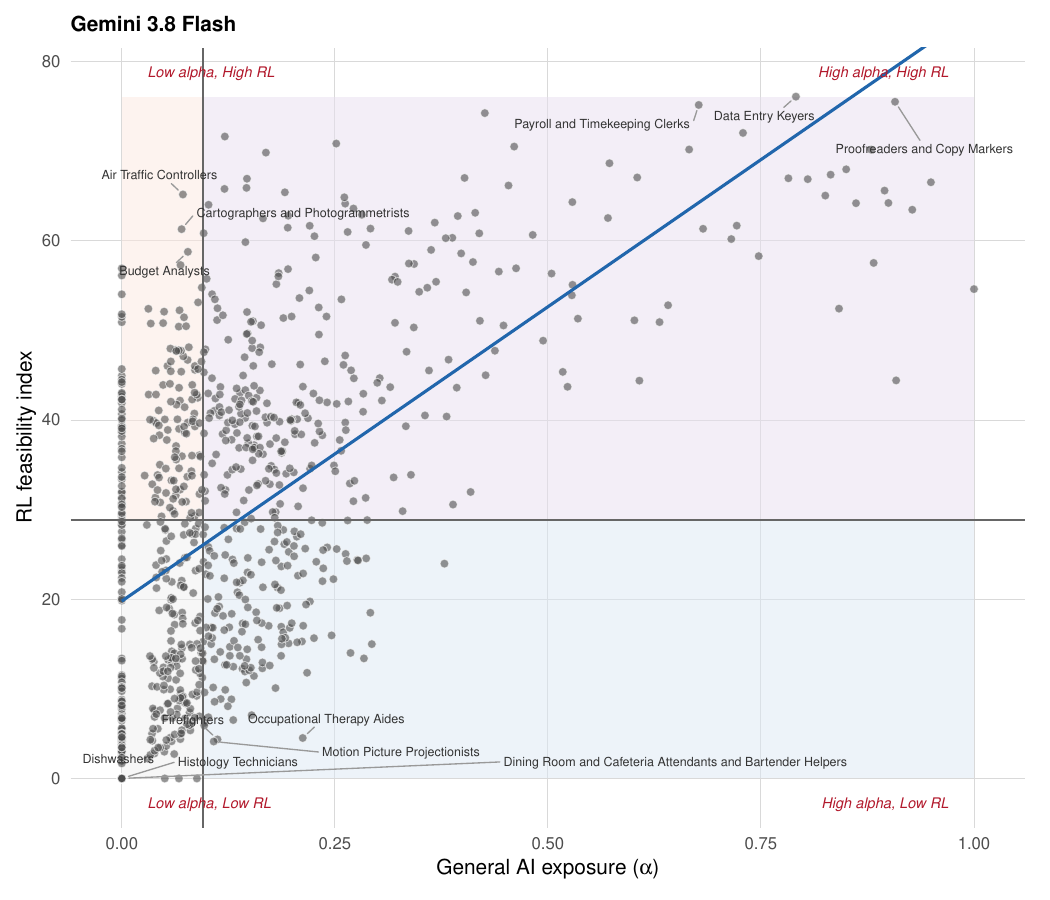}
    \caption*{\footnotesize \textit{Notes:} As Figure~\ref{fig:quadrant}, with the $\alpha$ measure of \citet{eloundou2024gpts} (importance-weighted share of E1 tasks, GPT-4 labels) on the horizontal axis. Each point is one occupation ($N = 894$). Solid lines mark sample medians; shaded regions denote the four quadrants defined by above/below-median splits on each axis. The three occupations furthest from the medians in each quadrant are labeled. Both indices are aggregated to occupations using O*NET 30.0 importance weights.}
\end{figure}

\begin{figure}[!htbp]
    \centering
    \caption{Occupation-level RL feasibility vs.\ general AI exposure ($\gamma$).}
    \label{fig:quadrant_gamma}
    \includegraphics[width=0.8\textwidth]{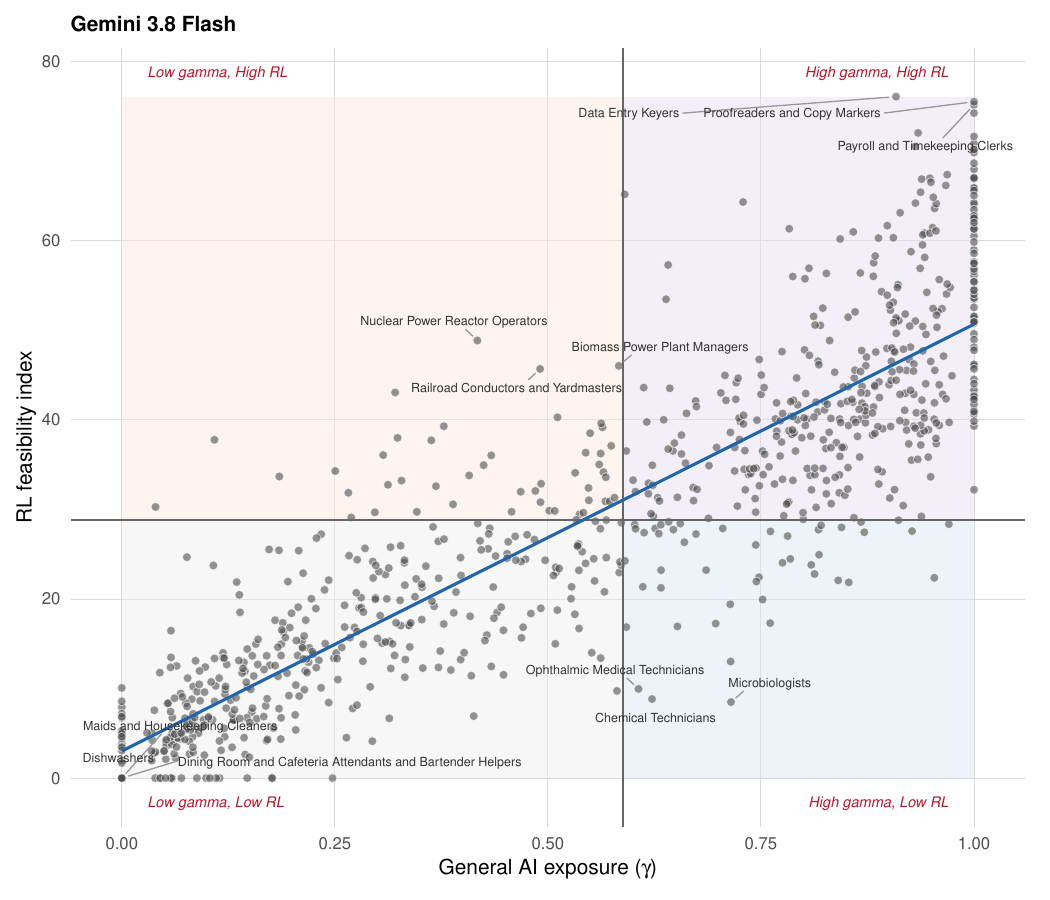}
    \caption*{\footnotesize \textit{Notes:} As Figure~\ref{fig:quadrant}, with the $\gamma$ measure of \citet{eloundou2024gpts} (E1 and E2 tasks at full weight, GPT-4 labels) on the horizontal axis. Each point is one occupation ($N = 894$). Solid lines mark sample medians; shaded regions denote the four quadrants defined by above/below-median splits on each axis. The three occupations furthest from the medians in each quadrant are labeled. Both indices are aggregated to occupations using O*NET 30.0 importance weights.}
\end{figure}

\begin{figure}[!htbp]
    \centering
    \caption{Occupation-level RL feasibility vs.\ general AI exposure (human-annotated $\beta$).}
    \label{fig:quadrant_beta_human}
    \includegraphics[width=0.8\textwidth]{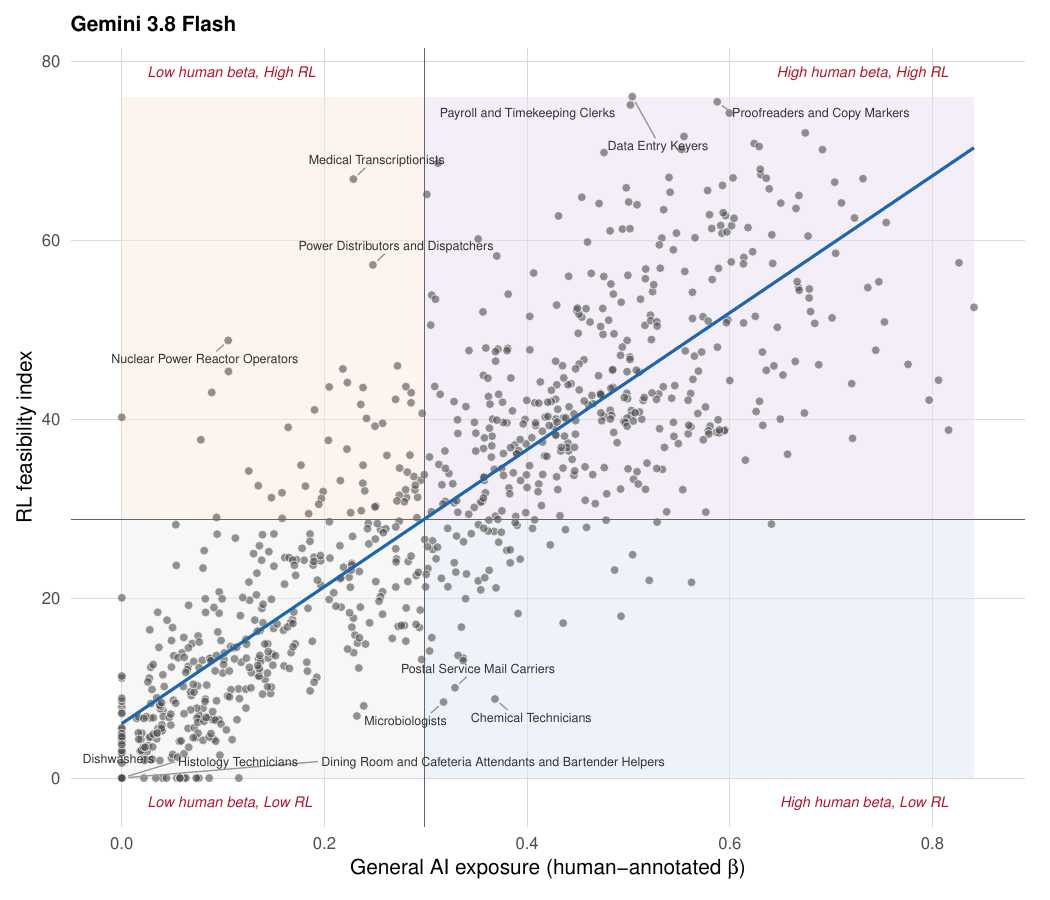}
    \caption*{\footnotesize \textit{Notes:} As Figure~\ref{fig:quadrant}, with the $\beta$ measure of \citet{eloundou2024gpts} computed from the human labels (E1 tasks at full weight, E2 tasks at half weight) on the horizontal axis. Each point is one occupation ($N = 894$). Solid lines mark sample medians; shaded regions denote the four quadrants defined by above/below-median splits on each axis. The three occupations furthest from the medians in each quadrant are labeled. Both indices are aggregated to occupations using O*NET 30.0 importance weights.}
\end{figure}

Figures~\ref{fig:quadrant_alpha} to \ref{fig:quadrant_beta_human} repeat Figure~\ref{fig:quadrant} for $\alpha$, $\gamma$, and human-annotated $\beta$, with medians recomputed for each measure. The $\gamma$ plot assigns 848 of the 894 occupations to the same quadrant as the $\beta$ plot, and the human-annotated $\beta$ plot assigns 786. The off-diagonal quadrants therefore contain largely the same occupations under these three measures. Nuclear power reactor operators, railroad conductors, and biomass power plant managers are among the labeled occupations with low general exposure and high RL feasibility, and microbiologists and chemical technicians are among those with the reverse profile. Two features of the plots are specific to one measure. Under $\gamma$, 80 occupations reach the maximum of 1, which produces the column of points at the right edge of Figure~\ref{fig:quadrant_gamma}. Under human-annotated $\beta$, medical transcriptionists move into the low-exposure, high-RL quadrant, because the human annotators rate their exposure at 0.23 against 0.87 for GPT-4.

The $\alpha$ plot differs most, assigning 602 occupations to the same quadrant as the $\beta$ plot. Because $\alpha$ counts only tasks an LLM can perform on its own, 183 occupations score exactly zero and the median is 0.095. The low-exposure, high-RL quadrant then fills with digital occupations whose tasks GPT-4 labels as exposed only with additional software, such as graphic designers, actuaries, cost estimators, and market research analysts, all with $\alpha = 0$ and RL scores between 44 and 54. The high-exposure, low-RL quadrant in turn includes physical occupations such as firefighters and occupational therapy aides, whose $\alpha$ scores of 0.11 and 0.21 sit just above the median while their RL scores are below 5.

\begin{figure}[!htbp]
    \centering
    \caption{Occupation-level RL feasibility vs.\ general AI exposure ($\beta$), gate-passing tasks.}
    \label{fig:quadrant_beta_gated}
    \includegraphics[width=0.8\textwidth]{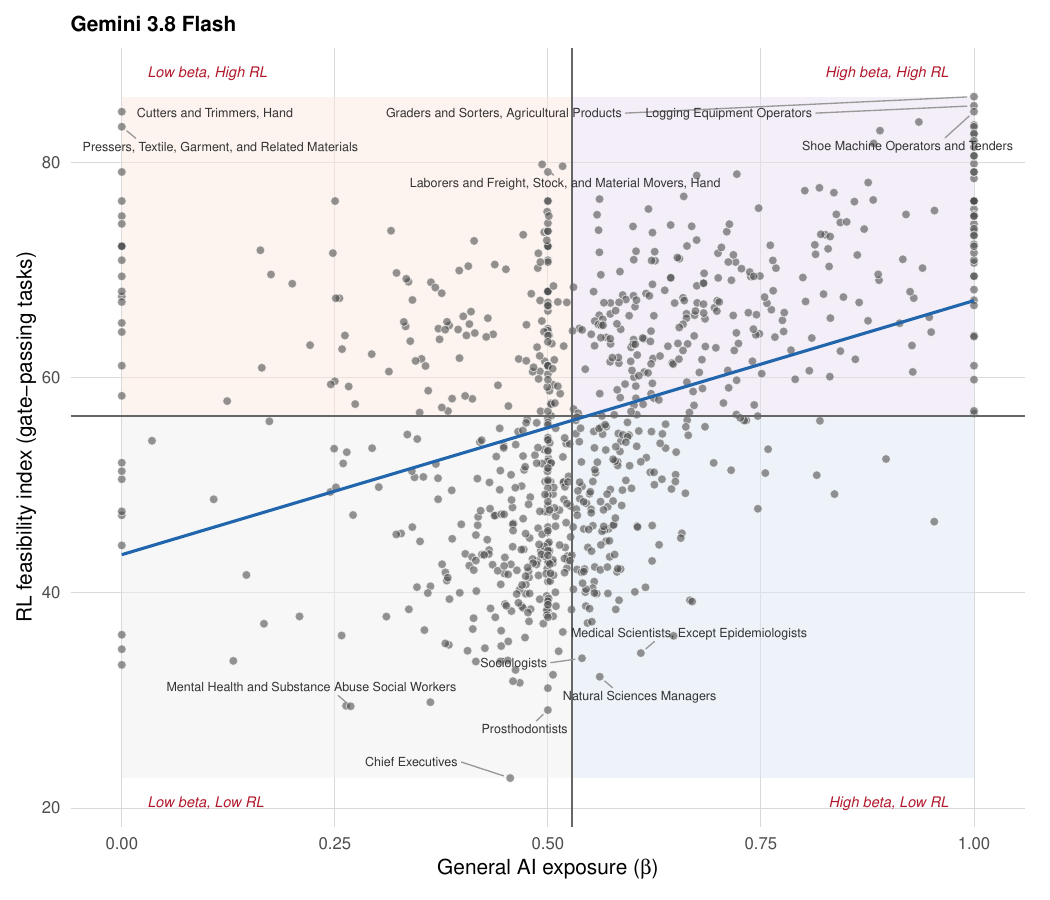}
    \caption*{\footnotesize \textit{Notes:} As Figure~\ref{fig:quadrant}, with the $\beta$ measure of \citet{eloundou2024gpts} (E1 tasks at full weight, E2 tasks at half weight, GPT-4 labels) on the horizontal axis. Both indices are re-aggregated over the 10,011 matched tasks with valid Gemini 3.8 Flash scores that pass the physical feasibility gate. Each point is one occupation ($N = 847$). Solid lines mark sample medians, recomputed for this sample; shaded regions denote the four quadrants defined by above/below-median splits on each axis. The three occupations furthest from the medians in each quadrant are labeled. Both indices are aggregated to occupations using O*NET 30.0 importance weights.}
\end{figure}

\begin{figure}[!htbp]
    \centering
    \caption{Occupation-level RL feasibility vs.\ general AI exposure ($\alpha$), gate-passing tasks.}
    \label{fig:quadrant_alpha_gated}
    \includegraphics[width=0.8\textwidth]{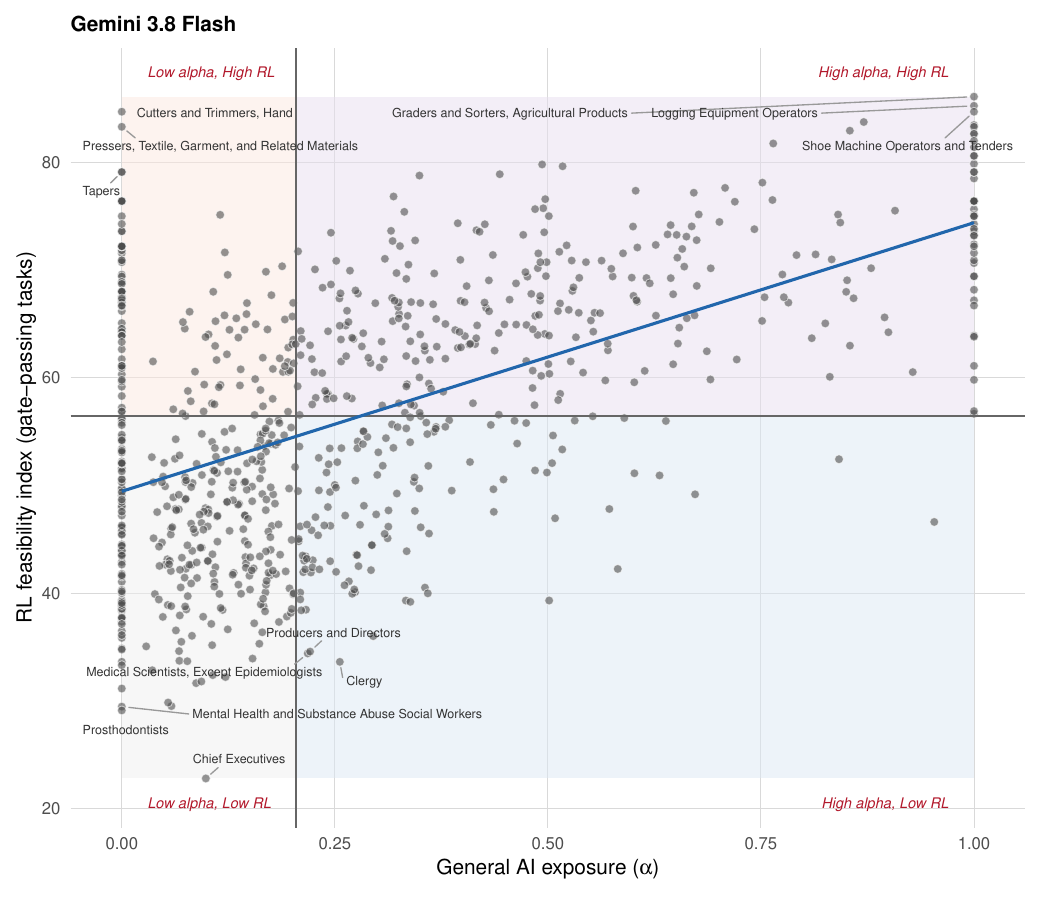}
    \caption*{\footnotesize \textit{Notes:} As Figure~\ref{fig:quadrant}, with the $\alpha$ measure of \citet{eloundou2024gpts} (importance-weighted share of E1 tasks, GPT-4 labels) on the horizontal axis. Both indices are re-aggregated over the 10,011 matched tasks with valid Gemini 3.8 Flash scores that pass the physical feasibility gate. Each point is one occupation ($N = 847$). Solid lines mark sample medians, recomputed for this sample; shaded regions denote the four quadrants defined by above/below-median splits on each axis. The three occupations furthest from the medians in each quadrant are labeled. Both indices are aggregated to occupations using O*NET 30.0 importance weights.}
\end{figure}

\begin{figure}[!htbp]
    \centering
    \caption{Occupation-level RL feasibility vs.\ general AI exposure ($\gamma$), gate-passing tasks.}
    \label{fig:quadrant_gamma_gated}
    \includegraphics[width=0.8\textwidth]{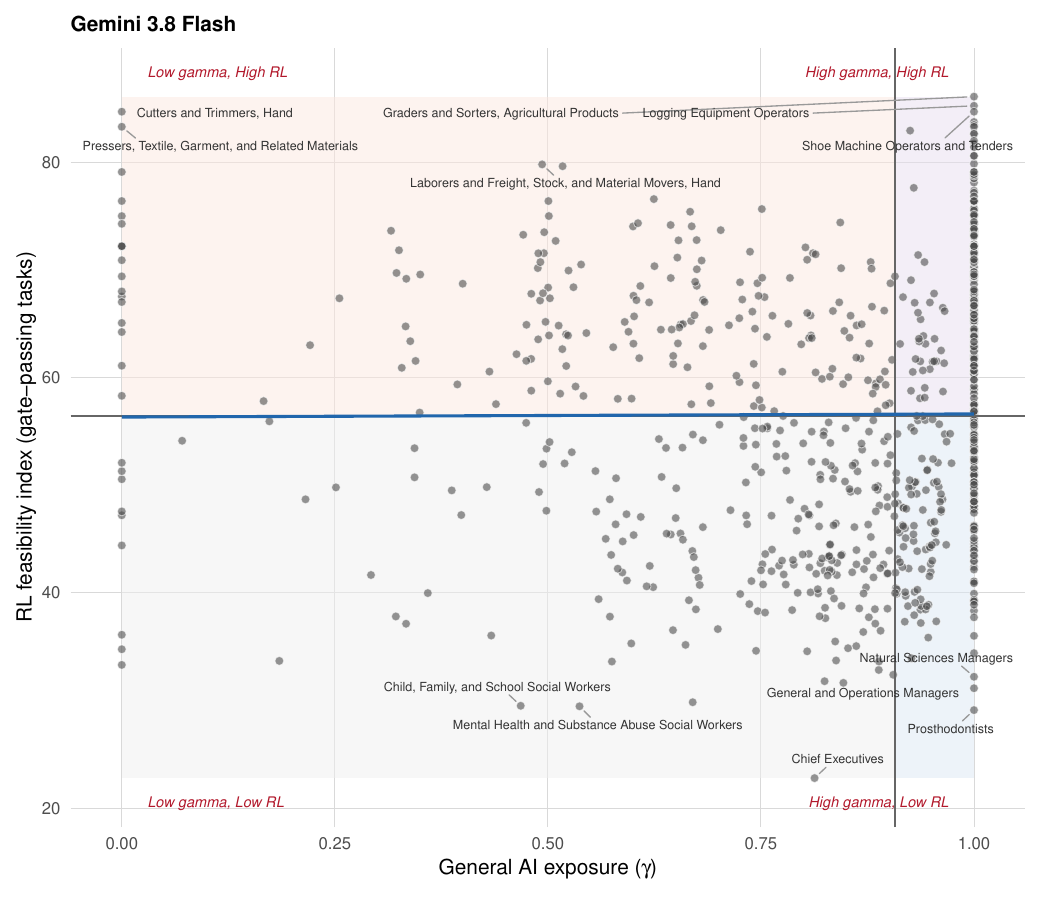}
    \caption*{\footnotesize \textit{Notes:} As Figure~\ref{fig:quadrant}, with the $\gamma$ measure of \citet{eloundou2024gpts} (E1 and E2 tasks at full weight, GPT-4 labels) on the horizontal axis. Both indices are re-aggregated over the 10,011 matched tasks with valid Gemini 3.8 Flash scores that pass the physical feasibility gate. Each point is one occupation ($N = 847$). Solid lines mark sample medians, recomputed for this sample; shaded regions denote the four quadrants defined by above/below-median splits on each axis. The three occupations furthest from the medians in each quadrant are labeled. Both indices are aggregated to occupations using O*NET 30.0 importance weights.}
\end{figure}

\begin{figure}[!htbp]
    \centering
    \caption{Occupation-level RL feasibility vs.\ general AI exposure (human-annotated $\beta$), gate-passing tasks.}
    \label{fig:quadrant_beta_human_gated}
    \includegraphics[width=0.8\textwidth]{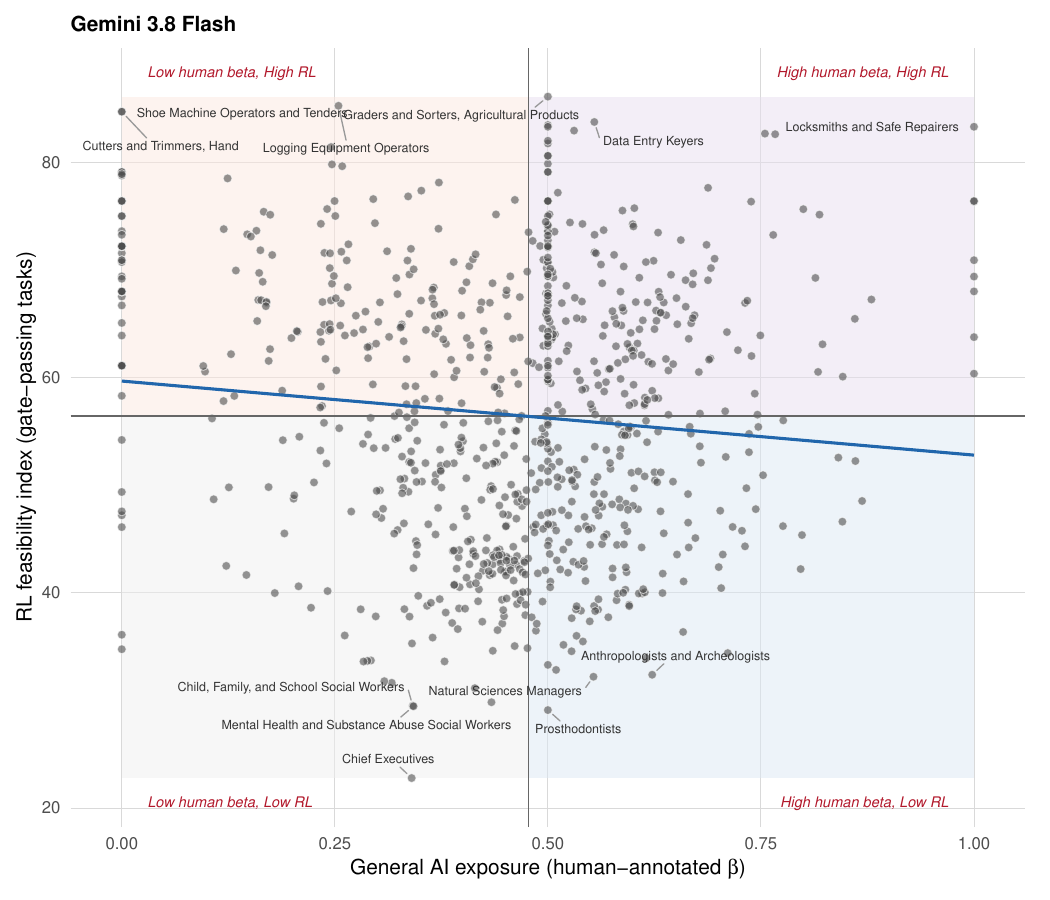}
    \caption*{\footnotesize \textit{Notes:} As Figure~\ref{fig:quadrant}, with the $\beta$ measure of \citet{eloundou2024gpts} computed from the human labels (E1 tasks at full weight, E2 tasks at half weight) on the horizontal axis. Both indices are re-aggregated over the 10,011 matched tasks with valid Gemini 3.8 Flash scores that pass the physical feasibility gate. Each point is one occupation ($N = 847$). Solid lines mark sample medians, recomputed for this sample; shaded regions denote the four quadrants defined by above/below-median splits on each axis. The three occupations furthest from the medians in each quadrant are labeled. Both indices are aggregated to occupations using O*NET 30.0 importance weights.}
\end{figure}

\clearpage
Figures~\ref{fig:quadrant_beta_gated} to \ref{fig:quadrant_beta_human_gated} repeat the four plots with both indices re-aggregated over the 10,011 matched tasks that pass the physical feasibility gate, which leaves 847 occupations, and with medians recomputed for this sample. The RL median rises from 28.9 to 56.4 because gate-failing tasks no longer contribute zeros, and the $\beta$ median rises from 0.36 to 0.53. Quadrant membership changes for most occupations: only 17 of the 150 occupations in the low-exposure, high-RL quadrant and 16 of the 150 in the high-exposure, low-RL quadrant occupied the same quadrant in the full sample. The $\alpha$ plot assigns 667 of the 847 occupations to the same quadrant as the $\beta$ plot, the $\gamma$ plot 619, and the human-annotated $\beta$ plot 573. Fifty-one occupations sit in the low-exposure, high-RL quadrant under all four measures and 35 in the high-exposure, low-RL quadrant. Under $\gamma$, 294 occupations reach the maximum of 1 and the median is 0.91, so the vertical median line sits close to the right edge of Figure~\ref{fig:quadrant_gamma_gated}. Under human-annotated $\beta$, the fitted line slopes downward, in line with the correlation of $-0.102$ in Table~\ref{tab:eloundou_alternatives}.

\clearpage
Table~\ref{tab:quadrant_descriptives} describes the workers in each quadrant of Figure~\ref{fig:quadrant}. We assign each occupation to a quadrant by the median splits of Figure~\ref{fig:quadrant} and pool the Revelio Labs position records active on 1 October 2022 (the pre-ChatGPT sample of Appendix~\ref{app:pre_chatgpt}) over the occupations in each quadrant. Section~\ref{sec:comparison} discusses the results.

\begin{table}[!htbp]
\centering
\caption{Workers by quadrant of RL feasibility and $\beta$, 1 October 2022}
\label{tab:quadrant_descriptives}
\scriptsize
\begin{tabular}{lrrrrr}
\toprule
\textbf{Statistic} & \textbf{Low $\beta$, low RL} & \textbf{High $\beta$, low RL} & \textbf{Low $\beta$, high RL} & \textbf{High $\beta$, high RL} & \textbf{All} \\
\midrule
Occupations & 393 & 48 & 48 & 399 & 888 \\
Positions (millions) & 15.0 & 5.1 & 3.1 & 70.4 & 93.7 \\
Share of positions (\%) & 16.1 & 5.4 & 3.3 & 75.2 & 100.0 \\
Share, Revelio-weighted (\%) & 17.0 & 5.4 & 3.4 & 74.2 & 100.0 \\
Mean salary (USD) & 61,610 & 122,970 & 75,798 & 99,348 & 93,783 \\
Median salary (USD) & 48,774 & 103,323 & 64,264 & 82,888 & 76,054 \\
Mean seniority (1--7) & 1.65 & 3.55 & 2.29 & 2.78 & 2.62 \\
Manager or above (\%) & 4.5 & 44.2 & 11.5 & 19.8 & 18.4 \\
Nonmetropolitan area (\%) & 24.1 & 18.5 & 23.0 & 18.2 & 19.3 \\
Ten largest metro areas (\%) & 25.8 & 34.9 & 27.5 & 32.9 & 31.7 \\
\bottomrule
\end{tabular}
\caption*{\footnotesize \textit{Notes:} Quadrants split occupations at the sample medians of the RL feasibility index and of $\beta$ (all tasks), as in Figure~\ref{fig:quadrant}; ``All'' pools the four quadrants. Statistics over Revelio Labs position records active on 1 October 2022 with a valid salary and an O*NET code among the 894 indexed occupations (93,698,554 records). Revelio-weighted shares apply Revelio's sampling weights. Seniority is Revelio's seven-level scale (1 = entry-level, 7 = executive); manager or above is levels 5 to 7. Metro-area shares are among the 92.9\% of records with a known metro area; the ten largest metro areas are ranked over all occupations.}
\end{table}

\clearpage
\section{Pre-ChatGPT Wage, Seniority, and Industry Gradients}
\label{app:pre_chatgpt}

This appendix reproduces the wage, seniority, and industry analyses from Section~\ref{sec:labour_market} using positions active on 1 October 2022 (before ChatGPT's release on 30 November 2022). The patterns are nearly identical to the recent sample, suggesting that the gradients were already present before ChatGPT's release.

\begin{figure}[t!]
    \centering
    \caption{Mean RL Feasibility Index (bars) by wage decile (1 October 2022, pre-ChatGPT).}
    \label{fig:wage_pre}
    \includegraphics[width=0.8\textwidth]{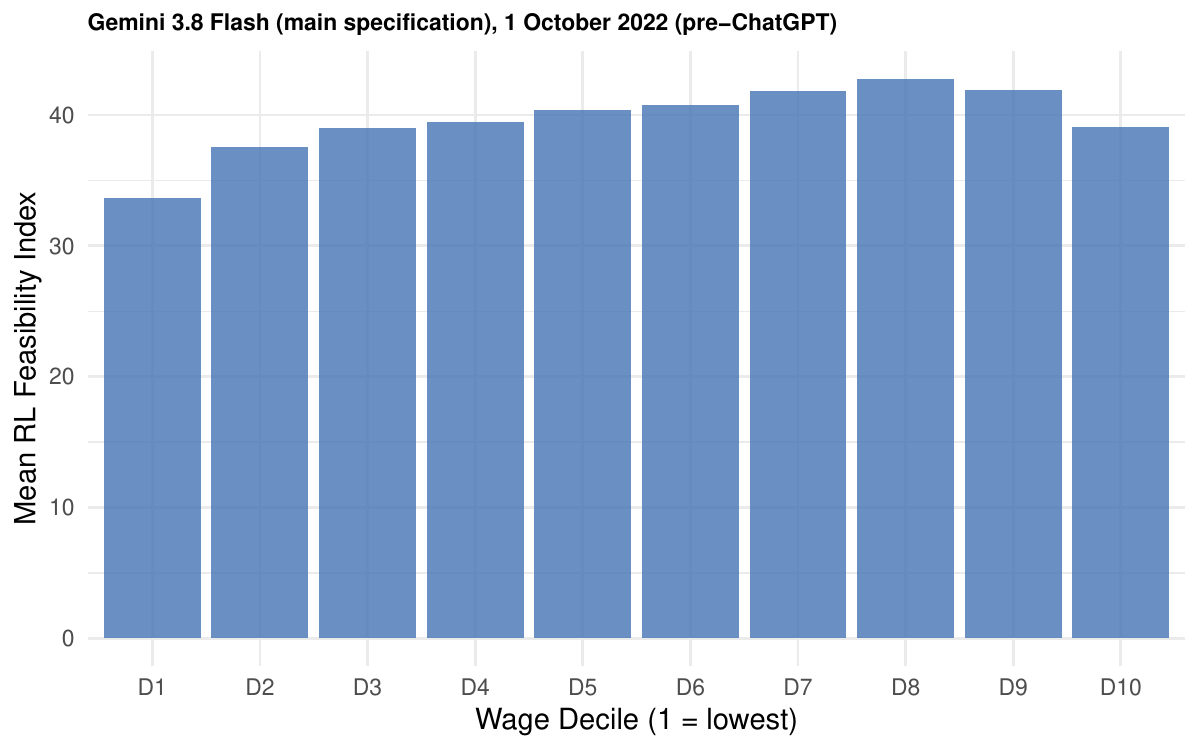}
    \caption*{\footnotesize \textit{Notes:} Bars show the employment-weighted mean RL Feasibility Index. Decile 1 is lowest-paid. Wage deciles are constructed from individual position salaries using Revelio Labs position records active on 1 October 2022 (93.7 million records), before ChatGPT's release. The matched sample covers 888 occupations.}
\end{figure}

\begin{figure}[t!]
    \centering
    \caption{Mean RL Feasibility Index (bars) by seniority level (1 October 2022, pre-ChatGPT).}
    \label{fig:seniority_pre}
    \includegraphics[width=0.8\textwidth]{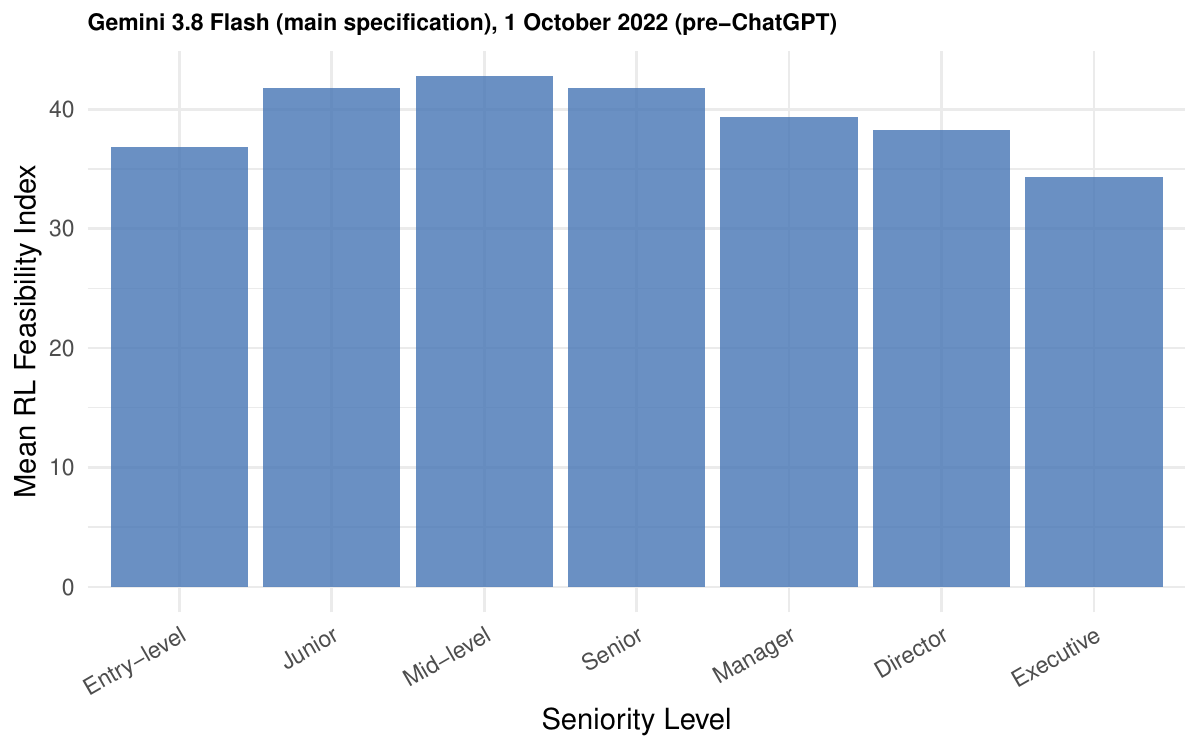}
    \caption*{\footnotesize \textit{Notes:} Bars show the employment-weighted mean RL Feasibility Index. Seniority levels range from 1 (entry-level) to 7 (executive). Computed from Revelio Labs position records active on 1 October 2022 (93.7 million records). The matched sample covers 888 occupations.}
\end{figure}

\begin{figure}[t!]
    \centering
    \caption{Mean RL Feasibility Index by NAICS sector (1 October 2022, pre-ChatGPT).}
    \label{fig:industry_pre}
    \includegraphics[width=0.8\textwidth]{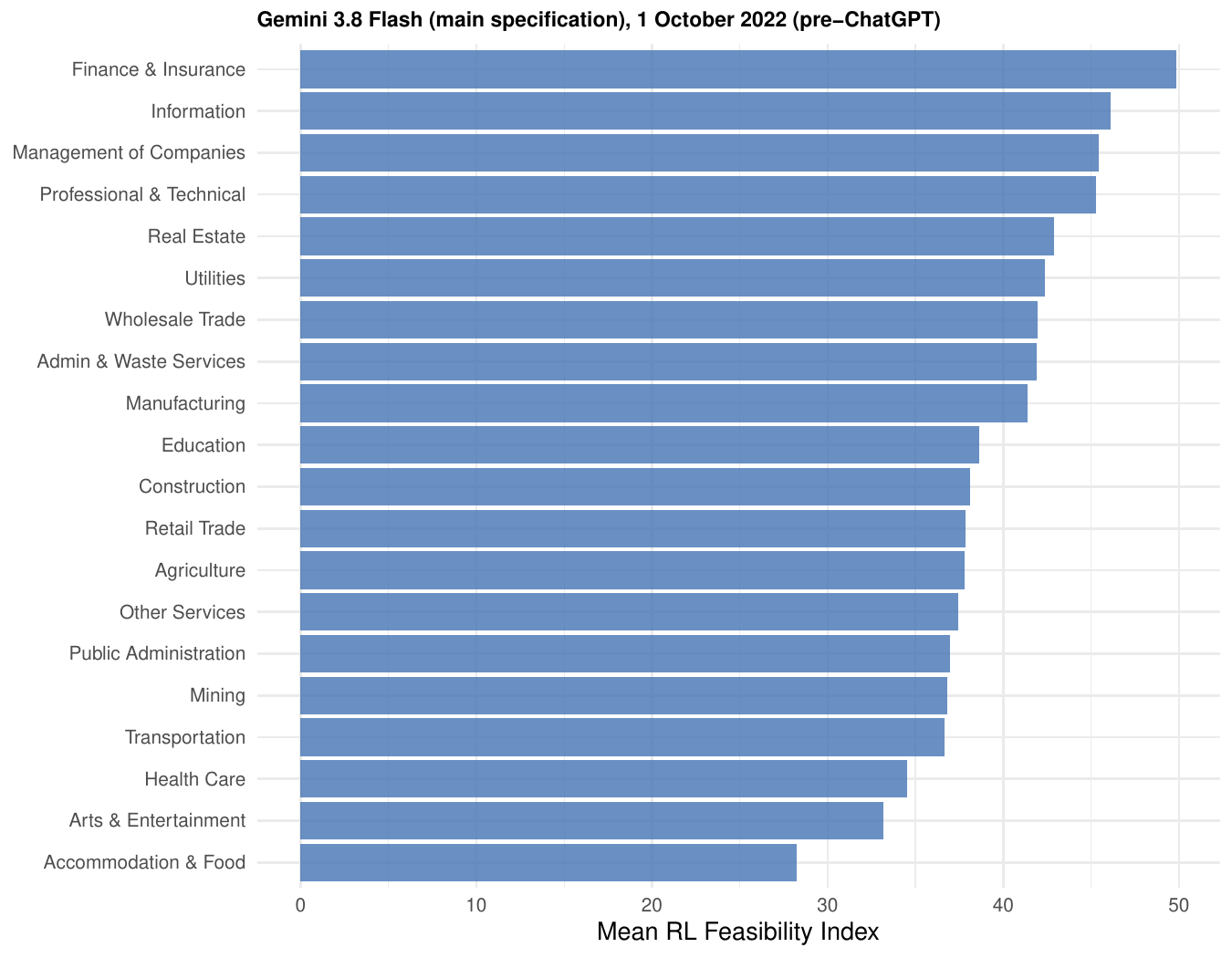}
    \caption*{\footnotesize \textit{Notes:} Bars show the employment-weighted mean RL Feasibility Index. Positions without a matched industry are excluded. Industries are 2-digit NAICS sectors. Computed from Revelio Labs position records active on 1 October 2022 (93.7 million records). The matched sample covers 888 occupations.}
\end{figure}

\begin{table}[!htbp] \centering 
  \caption{Occupation-level regressions of RL feasibility on wage and seniority (1 October 2022 (pre-ChatGPT)). Model: Gemini 3.8 Flash} 
  \label{tab:regression_pre_chatgpt} 
\scriptsize 
\begin{tabular}{@{\extracolsep{5pt}}lcc} 
\\[-1.8ex]\hline 
\hline \\[-1.8ex] 
\\[-1.8ex] & \multicolumn{2}{c}{RL Feasibility Index} \\ 
 & OLS & SOC major FE \\ 
\hline \\[-1.8ex] 
 Log(mean salary) & $-$1.473 & 3.056 \\ 
  & (2.679) & (2.306) \\ 
  & & \\ 
 Mean seniority & 34.780$^{***}$ & 20.315$^{***}$ \\ 
  & (4.162) & (2.991) \\ 
  & & \\ 
 Mean seniority squared & $-$4.906$^{***}$ & $-$3.468$^{***}$ \\ 
  & (0.731) & (0.519) \\ 
  & & \\ 
 Constant & $-$3.276 &  \\ 
  & (26.677) &  \\ 
  & & \\ 
\hline \\[-1.8ex] 
SOC major group FE & No & Yes \\ 
Observations & 888 & 888 \\ 
R$^{2}$ & 0.224 & 0.666 \\ 
Adjusted R$^{2}$ & 0.221 & 0.657 \\ 
\hline 
\hline \\[-1.8ex] 
\textit{Note:}  & \multicolumn{2}{r}{$^{*}$p$<$0.1; $^{**}$p$<$0.05; $^{***}$p$<$0.01} \\ 
\end{tabular} 
\caption*{\footnotesize \textit{Notes:} OLS and SOC-major-group fixed-effects regressions of the RL feasibility index of model Gemini 3.8 Flash on log mean salary and a quadratic in mean seniority. Unit of observation is an O*NET occupation ($N = 888$). Salary and seniority are computed from Revelio Labs position records active in 1 October 2022 (pre-ChatGPT). Standard errors in parentheses. $^{*}p<0.1$; $^{**}p<0.05$; $^{***}p<0.01$.}
\end{table}

\end{document}